\documentclass[aps,prx,twocolumn,superscriptaddress]{revtex4-2}

\usepackage[T1]{fontenc}
\usepackage[utf8]{inputenc}
\usepackage{lmodern}
\usepackage{microtype}

\usepackage{amsmath,amssymb,amsfonts,bm,mathtools}
\usepackage{mathrsfs}
\usepackage{braket}
\usepackage{dsfont}

\usepackage{graphicx}
\usepackage{xcolor}
\usepackage{subfigure}
\usepackage{tikz}
\usetikzlibrary{calc}
\usetikzlibrary{shapes.multipart}

\usepackage{booktabs,array}

\usepackage{float,verbatim,tabularx,multirow,appendix,wasysym}
\usepackage{amsthm}
\usepackage{textcomp}
\usepackage{yfonts}
\usepackage[normalem]{ulem}
\usepackage[colorlinks=true,linkcolor=blue,citecolor=blue,urlcolor=blue]{hyperref}

\newcommand{\e}{\mathrm{e}}
\newcommand{\Tr}{\operatorname{Tr}}

\newcommand{\mbb}[1]{\mathbb{#1}}

\renewcommand{\vec}[1]{{\mathbf #1}}

\newcommand{\avg}[2]{\left\langle #1\right\rangle_{#2}}

\newcommand{\Renyi}{R\'enyi}

\newcommand{\Order}[1]{\mathcal O\!\left(#1\right)}
\newcommand{\gc}{g_{\mathrm c}}

\newcommand{\FO}{\mathrm{FO}}
\newcommand{\WFO}{\mathrm{WFO}}

\newcommand{\stkout}[1]{\ifmmode\text{\sout{\ensuremath{#1}}}\else\sout{#1}\fi}

\begin{document}
\title{Finite-size scaling theory of Rényi thermal entropy for second-order, first-order, and weakly first-order quantum phase transitions}

\author{Zhe Wang}
\email{wangzhe90@westlake.edu.cn}
\affiliation{Department of Physics, School of Science and Research Center for Industries of the Future, Westlake University, Hangzhou 310030,  China}
\affiliation{Institute of Natural Sciences, Westlake Institute for Advanced Study, Hangzhou 310024, China}

\author{Yanzhang Zhu}
\affiliation{Department of Physics, School of Science and Research Center for Industries of the Future, Westlake University, Hangzhou 310030,  China}
\affiliation{Institute of Natural Sciences, Westlake Institute for Advanced Study, Hangzhou 310024, China}
\affiliation{State Key Laboratory of Surface Physics and Department of Physics, Fudan University, Shanghai 200433, China}

\author{Chunhao Guo }
\email{guochunhao@westlake.edu.cn}
\affiliation{Department of Physics, School of Science and Research Center for Industries of the Future, Westlake University, Hangzhou 310030, China}
\affiliation{Institute of Natural Sciences, Westlake Institute for Advanced Study, Hangzhou 310024, China}

\author{Bin-Bin Chen}
\affiliation{Peng Huanwu Collaborative Center for Research and Education, Beihang University, Beijing 100191, China}

\author{Yi-Ming Ding}
\affiliation{Department of Physics, School of Science and Research Center for Industries of the Future, Westlake University, Hangzhou 310030,  China}
\affiliation{Institute of Natural Sciences, Westlake Institute for Advanced Study, Hangzhou 310024, China}
\affiliation{State Key Laboratory of Surface Physics and Department of Physics, Fudan University, Shanghai 200433, China}

\author{Zenan Liu}
\affiliation{Department of Physics, South China University of Technology, Guangzhou 510640, China}

\author{Zheng Yan}
\email{zhengyan@westlake.edu.cn}
\affiliation{Department of Physics, School of Science and Research Center for Industries of the Future, Westlake University, Hangzhou 310030,  China}
\affiliation{Institute of Natural Sciences, Westlake Institute for Advanced Study, Hangzhou 310024, China}

\begin{abstract}
Establishing the nature of a quantum phase transition in finite-size simulations---whether continuous, first-order, or weakly first-order---is a fundamental challenge in quantum many-body computation.  The weakly first-order transition is particularly elusive: its anomalously large correlation length invariably mimics a continuous critical point at accessible system sizes, and no unified finite-size scaling theory has existed to distinguish these scenarios under realistic computational constraints.
In this work, we address this problem by introducing a finite-size scaling framework based on the R\'enyi thermal
entropy (RTE) and its derivative (DRTE).  We derive the finite-size scaling forms for the RTE and DRTE at continuous, first-order, and weakly first-order quantum phase transitions, and show that the DRTE isolates the universal contribution of the free energy and strengthens the characteristics of various phase transitions in finite sizes.  Using quantum Monte Carlo simulations, we demonstrate accurate data collapse and extraction of critical exponents at (2+1)-dimensional O($N$) critical points. Most importantly, the DRTE provides a definitive
signature of weakly first-order transitions at visitable sizes---a zero-crossing and a double-peak structure---which we observe in the debated deconfined quantum criticality candidates such as the $J$--$Q$ and SU(8) $J_1$--$J_2$ models.  Our approach offers a general, order-parameter-agnostic,  and numerically efficient diagnostic tool, resolving long-standing ambiguities between continuous and weakly first-order scenarios in the finite size simulations.

\end{abstract}

\date{\today}
\maketitle

\section{Introduction} 

Quantum phase transitions, associated spontaneous symmetry breaking, and critical behaviors are central
concepts in condensed matter and statistical physics~\cite{sachdev1999quantum,wannier1987statistical,girvin2019modern}. The nature of a phase transition is determined by universal contribution in the free energy or its derivatives, as rigorously defined in standard textbooks: a first-order transition is characterized by a discontinuity in the first derivative of the free energy, while a continuous transition involves  higher-order derivatives. In principle, an exact analytical calculation of the partition function would allow unambiguous identification of the transition type through these signatures. However, only a handful of systems admit exact solutions, and the vast majority must be studied numerically. In practice, detecting such  features in numerical simulations is highly challenging, as it requires extremely high-precision data to resolve the behavior of derivatives—particularly in strongly correlated many-body systems.

A long-standing and fundamental challenge lies in identifying genuine deconfined quantum critical points (DQCP) beyond the Landau-Ginzburg-Wilson paradigm. This framework describes continuous transitions between incompatible symmetry-breaking phases, such as the antiferromagnetic (AF) N\'eel order—where spin symmetry is broken—and the valence-bond solid (VBS)—where lattice-rotation symmetry is broken~\cite{senthil2004deconfined,sandvik2007evidence,sandvik2010continuous}. Such continuous transitions exhibit unconventional critical behavior and may host emergent symmetries and fractionalized excitations~\cite{Nahum2015PRL,ma2018dynamical,shao2016quantum,wang2017deconfined,senthil2024deconfined,mao2025detecting,yang2025deconfined,liu2025edge}. If realized in nature, DQCP would represent a fundamentally new class of quantum critical phenomena beyond the Landau-Ginzburg-Wilson framework—a discovery of comparable significance to the establishment of the paradigm itself. However, in concrete model realizations, it remains controversial whether candidate DQCPs represent true continuous transitions or are instead weakly first-order. Notable examples include nonlinear sigma models with topological terms~\cite{ma2020theory,lu2023nonlinear,wang2021phases,zhou2024mathrmso5}, the $J-Q$ models~\cite{sandvik2007evidence,zhao2022scaling,d2024entanglement,deng2024diagnosing,lou2009antiferromagnetic,wang2025probing}, and other bosonic or fermionic systems~\cite{Liu_2019,liaoGross2022,zhang2018continuous,liu2024deconfined}. The stakes are enormous: resolving this question determines whether we need an entirely new framework for quantum criticality or whether conventional Landau theory, augmented by proximate multicriticality, suffices.

The core difficulty stems from the fact that correlation lengths $\xi$ at these candidate DQCPs are often orders of magnitude longer than the system sizes accessible in simulations—even if the transition is weakly first-order rather than genuinely critical. In the regime $L \ll \xi$, a weakly first-order transition is nearly indistinguishable from a true continuous one. Consequently, the observed critical-like behavior closely mimics that of a continuous transition, making definitive discrimination extremely difficult. This phenomenon is understood within a physical picture in which the weakly first-order transition occurs in close proximity to a critical point—though some researchers argue that the underlying criticality is described by a complex conformal field theory (CFT) \cite{gorbenko2018walking,gorbenko2018walking2,zhou2024mathrmso5,tang2025boundary}, while others propose it arises from a multicritical point \cite{Jun2024SO(5),chen2024phases,liu2022emergence}. As a result, in finite-size simulations, weakly first-order transitions near a critical fixed point are nearly indistinguishable from continuous ones. Given current computational limits, reliably distinguishing between these scenarios across accessible system sizes remains a major challenge. 

The severity of this obstacle is best appreciated by surveying the existing diagnostic toolkit. Order parameters and Binder cumulants require delicate extrapolation to $L \to \infty$, but the extrapolation itself depends on assumptions about scaling corrections that are precisely what is debated, especially when the target value is close to zero. Entanglement entropy, one of the most sophisticated modern diagnostics, has recently been shown to be cut-dependent in ways not under theoretical control \cite{zhu2026bipartite,wang2026universal,zhu2026criticality}: several recent works studying the same $J$-$Q$ models reached the various conclusions from mutually incompatible signals depending on the lattice bipartition \cite{zhao2022scaling,deng2024diagnosing,d2024entanglement}. In other words, even the state-of-the-art diagnostic yields signals that change character depending on microscopic implementation choices. The common thread is that no existing method offers a binary, cut-independent, parameter-free signature of phase coexistence at weakly first-order transitions.

Crucially, there is currently no unified and effective finite-size scaling theory capable of robustly differentiating second-order and weakly first-order transitions. As a consequence, most numerical studies lack a definitive ``smoking-gun'' signature for identifying the true nature of the transition. In this work, we address this gap by developing a scaling theory based on the R\'enyi thermal entropy (RTE) and its derivative (DRTE)—physical observables that naturally cancel the leading analytic  background contributions to the free energy, thereby isolating the universal contribution that governs critical behavior. Unlike all existing diagnostics, the DRTE is defined on the full-system partition function $S^{(2)} = -\ln[Z(2\beta)/Z(\beta)^2]$—no subsystem, no cut, no bipartition dependence. Its signatures at a first-order transition (zero-crossing and double-peak structure) follow from thermodynamic identities: they are binary, qualitative features that exist if and only if two phases coexist, and are impossible at a genuine continuous transition. This requires no fitting, no extrapolation, and no prior knowledge of the order parameter. Moreover, the DRTE exhibits distinct and universal signatures in both first-order and weakly first-order transitions, enabling clear discrimination between different types of quantum phase transitions. In continuous and weakly first-order cases, it allows for the accurate extraction of critical exponents through data collapse, where the extracted exponents in the latter reflect those of the nearest critical fixed point.

We verify these predictions across a broad range of quantum critical systems. For continuous transitions, the RTE and DRTE reproduce the known critical points and correlation-length exponents of the $(2+1)$D O(1), O(2), and O(3) universality classes, while an independent thermal tensor-network calculation for the one-dimensional transverse-field Ising model further demonstrates that the framework is not restricted to quantum Monte Carlo (QMC). For strong first-order transitions, the anisotropic Heisenberg and checkerboard $J$-$Q$ models confirm the predicted zero crossing, opposite-sign double-peak structure, and first-order finite-size scaling. More importantly, applying the DRTE to the debated $J$-$Q_2$ and $J$-$Q_3$ models reveals first-order branch-competition signatures coexisting with pseudocritical scaling. In particular, for the $J$-$Q_2$ model, these characteristic weakly first-order signatures are already clearly resolved for system sizes $L=16$--$56$, precisely in a regime where conventional diagnostics remain strongly ambiguous. We further apply the DRTE to the SU(8) $J_1$-$J_2$ model, which recent entanglement-entropy scaling interpreted as exhibiting a continuous deconfined quantum critical point~\cite{song2025evolution}. In contrast, our DRTE results reveal a clear first-order component, providing new evidence that the  transition of the SU(8) $J_1$-$J_2$ model is weakly first-order. These results not only validate the scaling framework across established universality classes and first-order transitions, but also demonstrate its ability to uncover weakly first-order behavior in challenging DQCP candidates.

Taken together, our results establish the RTE and DRTE as a general, order-parameter-agnostic, bipartition-independent, and numerically efficient framework for characterizing quantum phase transitions. This provides, to our knowledge, the first qualitative resolution of the long-standing DQCP controversy using a definitive, parameter-free diagnostic at currently accessible simulation scales, and demonstrates the practical ability of the DRTE to distinguish continuous from weakly first-order behavior deep within the pseudocritical finite-size regime.

\section{Second order phase transition}

\subsection{Finite-size scaling form}
Let us begin with second-order quantum phase transitions.
For a quantum phase transition, the inverse temperature should
scale as $\beta = cL^z$, where $L$ is the linear system size,
$z$ is the dynamical critical exponent, and $c$ is a fixed
spacetime aspect ratio.  Since temperature is an irrelevant
scaling field, the free energy $F = -\frac{1}{\beta}\ln Z$
can be replaced by the logarithm of the partition function
$\ln Z$ for the purpose of extracting  universal contributions.

At a continuous transition the correlation length diverges,
and the free energy acquires a universal contribution.  According to
the finite-size scaling hypothesis~\cite{fisher1972scaling,
sandvik2010computational}, any scaling  observable $Q$  near
the critical point $g_c$ takes the scaling form
\begin{equation}
    Q(\Delta g, L, c) = L^\sigma \tilde{S}( \Delta g L^{1/\nu}, c),
    \label{Qterm}
\end{equation}
where $\Delta g = g - g_c$, $\nu$ is the correlation-length
exponent, $\sigma$ is a scaling power, and $\tilde{S}$ is a
universal scaling function.  The universal contribution of the free
energy is scale‑invariant ($\sigma = 0$)
\cite{Wilson1971Renormalization,fisher1972scaling}, hence
$\ln Z$ can be decomposed as
\begin{equation}
    \ln Z(g,\beta,L) = a(g)\beta L^{d}
    + \mathcal{A}_{\rm sub}(g,L,\beta)
    + \tilde{S}(\Delta g L^{1/\nu}, c).
    \label{lns}
\end{equation}
Here $a(g)\beta L^{d}$ is the leading volume‑law term and
$\mathcal{A}_{\rm sub}(g,L,\beta) =
\beta\bigl(a_1(g)L^{d-1} + a_2(g)L^{d-2} + \cdots\bigr)$
collects sub‑leading analytic background contributions.  The analytic background
terms dominate $\ln Z$ and obscure the  universal
part $\tilde{S}$, making direct numerical detection of
critical universalities challenging.

To isolate $\tilde{S}$, we should construct a quantity in which the
dominant analytic background contributions cancel.  For a quantum phase transition, temperature is an irrelevant
scaling field.  Along the finite‑size scaling trajectory
$\beta \sim L^z$, the analytic background is expected to be
linear in $\beta$ at leading order.  Evaluating
Eq.~\eqref{lns} at inverse temperatures $2\beta$ and $\beta$,
and multiplying the latter by two, we obtain
\begin{equation}
\begin{aligned}
\ln Z(2\beta) &= 2a(g)\beta L^{d}
                + 2\mathcal{A}_{\rm sub}(g,L,\beta)
                + \tilde{S}(\Delta g L^{1/\nu}, 2c), \\[2pt]
2\ln Z(\beta) &= 2a(g)\beta L^{d}
                + 2\mathcal{A}_{\rm sub}(g,L,\beta)
                + 2\tilde{S}(\Delta g L^{1/\nu}, c).
\end{aligned}
\label{eq:lnZ2beta}
\end{equation}
Then the analytic background cancels, leaving
\begin{equation}
 2\ln Z(\beta)-\ln Z(2\beta)
= 2\tilde{S}(\Delta g L^{1/\nu}, c)
  - \tilde{S}(\Delta g L^{1/\nu}, 2c).
\label{eq:differ}
\end{equation}
In the present work the aspect ratio is fixed to $c=1$, so
the right‑hand side depends only on the scaling variable
$x = \Delta g L^{1/\nu}$.  For convenience we define the
combined universal scaling function
\begin{equation}
\tilde{S}(\Delta g L^{1/\nu})
\equiv 2\tilde{S}(\Delta g L^{1/\nu}, c)
   - \tilde{S}(\Delta g L^{1/\nu}, 2c),
\end{equation}
evaluated at $c=1$, so that
\begin{equation}
 2\ln Z(\beta)-\ln Z(2\beta)
= \tilde{S}(\Delta g L^{1/\nu}).
\label{Eq:differ}
\end{equation}
This identifies the combination
$2\ln Z(\beta)-\ln Z(2\beta)$ as the quantity that isolates
the universal critical contribution.  In
Appendix~\ref{sec:continuous}, within the framework of a
low-energy effective theory, we prove---based on the
CFT tower---that this difference indeed captures only the
universal critical part of the free energy.

\begin{figure}[!tb]
\centering
 
\includegraphics[width=0.38\textwidth]{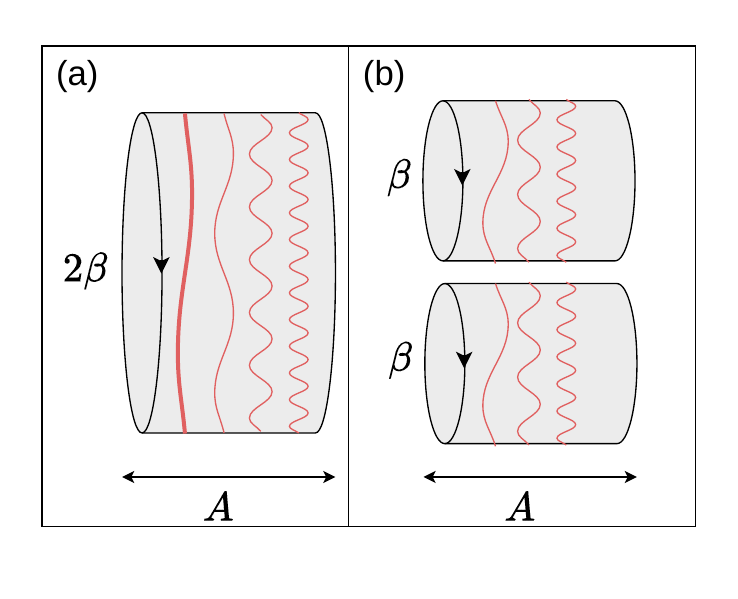}
\vspace{-20pt}
\caption{ A geometrical presentation of two partition functions  of  the whole system ($A$) (a) $Z(2\beta)=\mathrm{Tr}(e^{-2\beta H})$ and (b) $Z(\beta)=\mathrm{Tr}(e^{-\beta H})$, where $H$ is the Hamiltonian of the system. Near a quantum critical point with dynamical exponent \(z = 1\), the system possesses emergent Lorentz invariance, which puts the imaginary time \(\beta\) and the spatial size \(L\) on the same footing (\(\beta \propto L\)). The  Rényi thermal entropy  then compares the free energies of two Euclidean spacetime blocks whose temporal circumferences are \(2\beta\) and \(\beta\), respectively. In close analogy with the ordinary Casimir effect~\cite{Casimir1948Attraction,Plunien1986Casimir,Milton2001Casimir,FisherdeGennes1978,Krech1994,Cardy1988}, the finite imaginary-time circumference \(\beta\) acts as a “sieve” for critical fluctuations (different modes are depicted by red wavy lines of varying wavelengths): a larger circumference admits a broader modes, and the resulting difference in free energy can be understood as the energy cost of excluding long-wavelength (thick red wavy lines) modes when the temporal extent is reduced. We term this universal difference, which is entirely dominated by long-wavelength modes, the \textit{spacetime Casimir effect}.
}
\vspace{-10pt} 
\label{fig:sc}
\end{figure}

The above derivation shows that the difference
$ 2\ln Z(\beta)-\ln Z(2\beta)$ isolates the universal
part $\tilde{S}$ through an 
cancellation of the analytic background.  This result admits a transparent physical interpretation within the path-integral formalism: for dynamic exponent $z=1$, the system exhibits spacetime isotropy (emergent Lorentz invariance), placing imaginary time $\beta$ and spatial size $L$ on an equal footing. Specifically, Eq.~\eqref{Eq:differ} measures the
free‑energy difference between two Euclidean spacetime
manifolds whose temporal circumferences are $2\beta$ and
$\beta$ (see Fig.~\ref{fig:sc}).
In close analogy with the ordinary Casimir
effect~\cite{Casimir1948Attraction,Plunien1986Casimir,
Milton2001Casimir,FisherdeGennes1978,Krech1994,Cardy1988,
Ma2019Enhanced}, the finite temporal circumference of the
smaller block acts as a ``sieve'' for critical fluctuations:
a larger circumference accommodates a broader
modes, while tightening the temporal extent squeezes out soft
(long‑wavelength) modes at a cost in free energy (see
Fig.~\ref{fig:sc}).  We call this contribution the
\textit{spacetime Casimir effect}.  It is dominated by
infrared modes and yields an $O(1)$ universal scaling
function $\tilde{S}(\Delta g L^{1/\nu})$
(see Appendix~\ref{sec:casimireffect} for details).

From an information-theoretic perspective, the combination appearing in
Eq.~\eqref{eq:differ} is precisely the second‑order \Renyi{}
thermal entropy---also known as the purity---a fundamental
quantity in information
theory~\cite{renyi1961measures,Ozawa_2024,e24050706,
Amico2008entanglement,Eisert2010,Laflorencie2016,zeng2019}:
\begin{equation}
    S^{(2)} = -\ln\frac{Z(2\beta)}{Z(\beta)^2}.
\end{equation}
This definition extends to arbitrary order $n$ as
$S^{(n)} = \frac{1}{1-n}\ln[Z(n\beta)/Z(\beta)^n]$.
For simplicity, we focus on the second‑order R\'enyi thermal
entropy ($n=2$) throughout this work.

The above analysis shows that the RTE is a dimensionless
quantity dominated by long-wavelength fluctuations.
Its scaling dimension is therefore $\sigma=0$ in
Eq.~\eqref{Qterm}, and it inherits the scaling form
\begin{equation}
S^{(n)}(\Delta g, L) = \tilde{S}(\Delta g L^{1/\nu}).
\label{eq:rte}
\end{equation}
Differentiating Eq.~\eqref{eq:rte} yields the scaling
behavior of the derivative of the RTE (DRTE):
\begin{equation}
\frac{\partial S^{(n)}(\Delta g, L)}{\partial \Delta g}
= \tilde{S}'(\Delta g L^{1/\nu})\, L^{1/\nu},
\label{eq:drte}
\end{equation}
where $\tilde{S}'(x) = \partial \tilde{S}(x)/\partial x$
and the factor $L^{1/\nu}$ arises from the derivative of
the scaling variable $x = \Delta g L^{1/\nu}$ with respect
to $\Delta g$.

The above analysis implies two characteristic signatures:
(i) the RTE is dimensionless and therefore exhibits a
crossing at the critical point, analogous to the Binder
ratio; (ii) the DRTE allows the critical exponent $\nu$ to
be extracted through data collapse.  To verify these
predictions, we expand the scaling function as a polynomial,
$\tilde{S}(x) = \sum_{k=0}^{k_{\max}}
\frac{1}{k!}\tilde{S}^{(k)}(0)\, x^{k}$, and perform a
global fit to the RTE and DRTE data.  In the following, we
demonstrate perfect data collapse near the well‑established
(2+1)D O(1) ($Z_2$), O(2), and O(3) quantum critical
points.  The critical point $g_c$, the correlation‑length
exponent $\nu$, and the expansion coefficients
$\tilde{S}^{(k)}(0)$ are treated as fitting parameters; we
use $k_{\max} = 6$ throughout the numerical analysis.

\begin{figure*}[!htp]
\centering
\includegraphics[width=\textwidth]{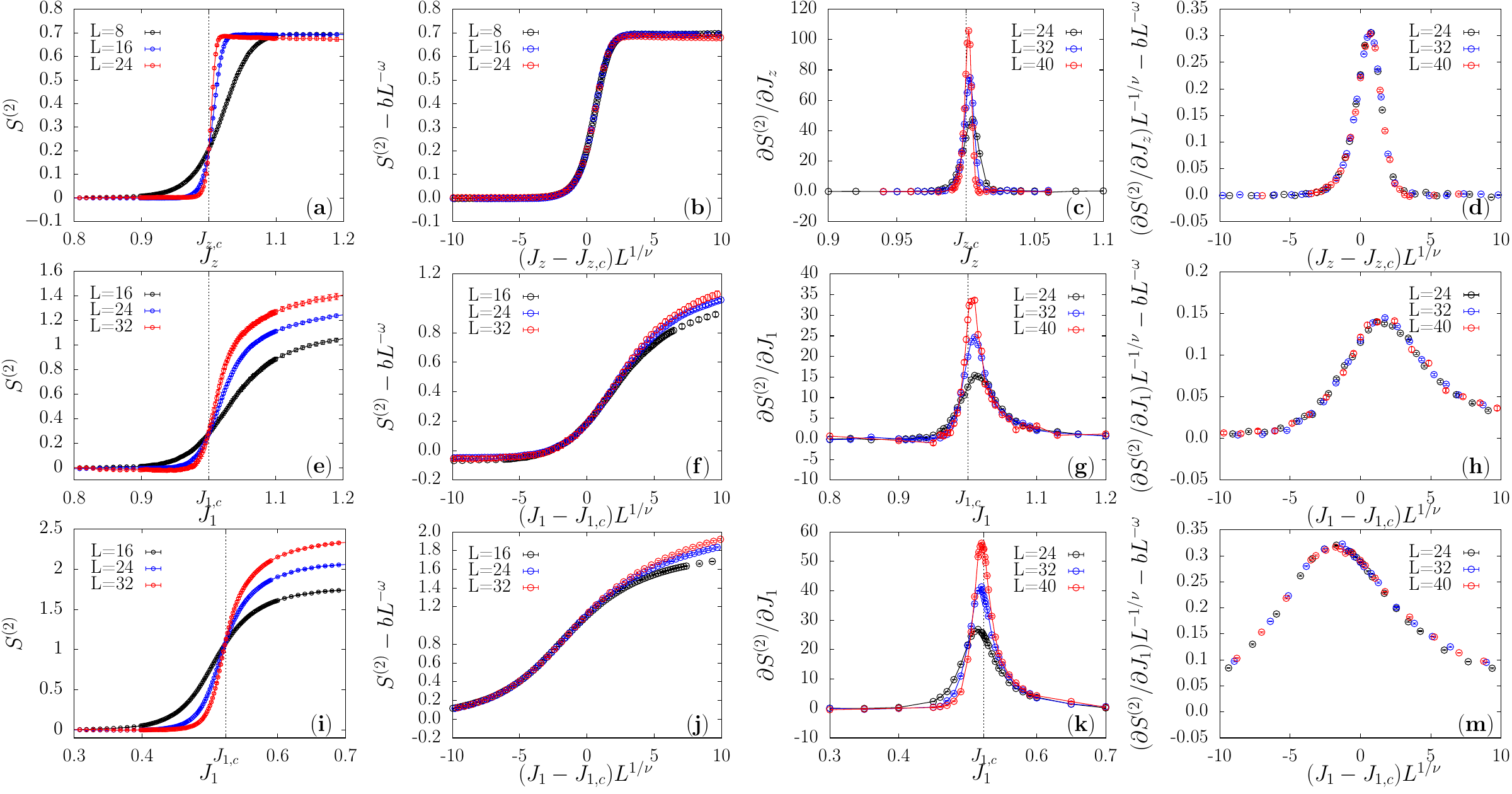}
\caption{(a,c) Second-order RTE and its derivative DRTE, and (b,d) corresponding data collapse analysis near the QCP of the bilayer Ising–Heisenberg model [(2+1)D Ising universality]. We set $J=3.045$ and take $J_{z}$ as the tuning parameter. (e--h) Results for the dimerized anisotropic Heisenberg model with $\Delta=0.9$ [(2+1)D O(2) universality], using $J_{2}=2.1035$ and tuning $ J_1 $.  (i--m) Results for the dimerized isotropic Heisenberg model with $\Delta=1$  [(2+1)D O(3) universality], using $J_{2}=1$ and tuning $ J_1 $. To achieve good data collapse, we include a subleading correction: $Q(g,L)=L^{k}[\tilde{S}((g-g_c)L^{1/\nu})+bL^{-\omega}]$. $\omega = 0.830$, $0.789$, and $0.782$ correspond to the (2+1)D Ising, O(2), and O(3) critical points, respectively~\cite{Andrea2002Critical,Campostrini2001Critical,Campostrini2002Critical,Deng2005Surface}.  For $Q=S^{(2)}$, $k=0$. $k=1/\nu$ when $Q$ is DRTE.}
\label{fig:qcp}
\end{figure*}

\subsection{Numerical methods}
The RTE can be expressed in terms of the ratio of partition functions at two different temperatures: $S^{(2)} =-\ln\frac{Z(2\beta)}{Z(\beta)^2} $.  We compute the partition functions using the bipartite reweight-annealing algorithm~\cite{wang2024ee,ding2024tracking,wang2026addressing}  in combination with  stochastic series expansion (SSE)  QMC simulations~\cite{Sandvik1999,sandvik2010computational,sandvik2019stochastic,Syljuaasen2002,yan2019sweeping,yan2020improved}. Following a method analogous to that used for the derivative of R\'enyi entanglement entropy and negativity~\cite{wang2024ee,wu2020entanglement}, the derivative of the RTE with respect to a general parameter $ g $ of the Hamiltonian $H$ is given by:
\begin{equation}
\frac{dS^{(n)}}{dg}=\frac{1}{1-n}\bigg[-n\beta\bigg\langle \frac{dH}{dg} \bigg\rangle_{Z(n\beta)}+n\beta\bigg\langle \frac{dH}{dg} \bigg\rangle_{Z(\beta)} \bigg]
\label{derivative}
\end{equation}
Here, the two expectation values are evaluated separately using ensembles corresponding to $ Z(n\beta) $ and $ Z(\beta) $, respectively. While the RTE is calculated using the reweight-annealing QMC method \cite{ding2024reweightannealing,dai2024residual,ma2024defining}, the DRTE is directly obtained from the two expectation values in Eq.~(\ref{derivative}); see Appendix \ref{sec:appm} for details. Considering the dynamical critical exponents  $ z = 1 $ of the models studied in this work, the $ \beta= L $ is set to approach the quantum critical regime.

\subsection{Ising transition} We study the bilayer Ising-Heisenberg model with a (2+1)D Ising QCP~\cite{wu2023classical}. The Hamiltonian is given by:
\begin{equation}
H=-J_{z}\sum_{\alpha=1,2}\sum_{\langle ij\rangle}S_{i,\alpha}^{z}S_{j,\alpha}^{z}+J\sum_{i}\vec{S}_{i,1}\cdot\vec{S}_{i,2},
\end{equation}
where $\langle ij\rangle$ denotes the nearest-neighbor interaction, and $\alpha=1,2$ labels the two layers. $-J_{z}<0$ represents the intralayer ferromagnetic (FM) Ising interaction, while $J>0$ represents the interlayer antiferromagnetic (AF) Heisenberg interaction. For $J/J_{z}\ll 1$, the ground state exhibits $\mbb{Z}_{2}$ symmetry-breaking Ising order. In the opposite limit,  $J/J_{z}\gg 1$, the system forms interlayer spin-singlet states.  The QCP separating  two phases is  at $J/J_{z}=3.045(2)$ and belongs to the (2+1)D Ising universality class~\cite{wu2023classical}.

We set $ J = 3.045 $ and use $ J_z $ as the tuning parameter, studying the second-order RTE and DRTE near the QCP at $ J_{z,c} = 1 $ for various lattice sizes. The results are shown in Figs.~\ref{fig:qcp}(a) and (c). The RTE curves for different system sizes intersect at the QCP, while the DRTE exhibits a clear peak at the critical point. We fit the scaling forms in Eqs.~(\ref{eq:rte}) and (\ref{eq:drte}) to the RTE and DRTE data, respectively. The fits yield $ J_{z,c} = 0.9999(1) $, $ \nu = 0.629(7) $ from RTE, and $ J_{z,c} = 1.000(3) $, $ \nu = 0.63(4) $ from DRTE. These values are in agreement with the known QCP $ J_{z,c} = 1 $ ~\cite{wu2023classical} and the (2+1)D Ising universality class value $ \nu = 0.63012 $ ~\cite{Guida1998,Deng2003Simultaneous,wu2023classical}. The resulting universal scaling functions $ \tilde{S}(\Delta g L^{1/\nu}) $ and $ \tilde{S}'(\Delta g L^{1/\nu}) $ with $\Delta g=J_z-J_{z,c}$, obtained from data collapse, are displayed in Figs.~\ref{fig:qcp}(b) and (d), respectively.

\subsection{Continuous symmetry phase transitions} We then investigate QCPs of the dimerized AF Heisenberg model~\cite{Matsumoto2001,ding2018engineering,zhu2022exotics}. The Hamiltonian reads:
\begin{equation}
H=J_{1}\sum_{\langle ij\rangle}D_{ij}+J_{2}\sum_{\langle ij\rangle'}D_{ij},
\label{anisotropicHeisenberg }
\end{equation}
in which $J_2 > J_1 > 0$ and $D_{ij}=S_{i}^{x}S_{j}^{x}+S_{i}^{y}S_{j}^{y}+\Delta S_{i}^{z}S_{j}^{z}$ represents the nearest-neighbor AF Heisenberg interaction with anisotropy parameter $\Delta$. In the regime where $J_{2}/J_{1}\simeq 1$, the ground state  exhibits  AF order. For $0<\Delta<1$, the AF order parameter lies in the $xy$-plane breaking the O(2) symmetry. For $\Delta>1$, the order parameter  aligns  along the $z$-axis, breaking the $\mbb{Z}_{2}$ spin flip symmetry. When  $\Delta=1$, the interaction is isotropic, and the AF order breaks the O(3) symmetry. In the limit $J_{2}/J_{1}\gg 1$, the ground state is the gapped dimer phase. The QCP from the gapped phase to the AF order belongs to the (2+1)D O(2)/Ising/O(3) universality class when $0<\Delta<1$/ $\Delta>1$ / $\Delta=1$ ~\cite{Matsumoto2001,ding2018engineering,zhu2022exotics}. For a zero-temperature system with ground-state degeneracy $ d_e $, the RTE actually reflects the logarithmic degeneracy:  $S^{(2)} =\ln d_e$. As shown in Fig.\ref{fig:qcp}, $S^{(2)}$ effectively probes both the degeneracy of discrete symmetry-breaking phases and the Anderson tower of states associated with continuous symmetry-breaking~\cite{deng2023improved,mao2023sampling,mao2025detecting}.

We evaluate the RTE and its derivative near the QCPs for both the easy-plane case ($\Delta=0.9$, O(2) criticality)  and the isotropic case ($\Delta=1$, O(3) criticality), with results shown in Figs.~\ref{fig:qcp}(e,g) and (i,k), respectively. In each case, $J_{2}$ is fixed and  $J_{1}$ is tuned across the transition, as detailed in the caption. Similar to the Ising  case, the RTE curves for different system sizes intersect at the QCPs, and their derivatives exhibit clear peaks at these points. We apply the scaling forms in Eqs.~(\ref{eq:rte}) and (\ref{eq:drte}) to data near the (2+1)D O(2) and O(3) QCPs. For the O(2) QCP, fitting the RTE yields $ J_{1,c} = 0.997(1) $, $ \nu = 0.67(2) $, while the DRTE gives $ J_{1,c} = 0.9994(8) $, $ \nu = 0.66(1) $. These values are consistent with the known critical parameters $ J_{1,c} = 1 $ and $ \nu = 0.6703 $~\cite{Guida1998,zhu2022exotics}.

For the O(3) QCP, fitting the RTE data yields $ J_{1,c} = 0.523(1) $, $ \nu = 0.703(7) $, while the DRTE gives $ J_{1,c} = 0.525(1) $, $ \nu = 0.708(8) $. These values are consistent with the known critical parameters $ J_{1,c} = 0.52337 $ and $ \nu = 0.7073 $~\cite{Guida1998,Matsumoto2001,ding2018engineering}. The universal scaling functions near the O(2) and O(3) QCPs, obtained from data collapse, are shown in Figs.~\ref{fig:qcp}(f,h) and (j,m), respectively.

In summary, all the data show that the RTE does have a crossing at a critical point and the DRTE does extract the critical exponent $\nu$ by data collapse, which is highly consistent with our theoretical analysis.

\section{First-order phase transition}
\subsection{Finite-size scaling form}

At a zero-temperature first-order transition, two thermodynamic branches have equal energy density at the transition but different generalized-force densities.  We write the finite-size energies of the two branches as
\begin{equation}
    E_\alpha(g,L)
    =L^d e_\alpha(g)+E_\alpha^{\rm sub}(g,L),
    \qquad \alpha=1,2,
    \label{eq:FObranchEnergies1}
\end{equation}
where \(e_\alpha(g)\) denotes the leading energy density of
branch \(\alpha\), while \(E_\alpha^{\rm sub}(g,L)\) collects
subextensive  corrections. At the transition point,
\begin{equation}
    e_1(\gc)=e_2(\gc),
    \qquad
    \ell=\partial_g[e_2(g)-e_1(g)]_{g=\gc}\ne0 .
    \label{eq:latentForce}
\end{equation}
Here,  $\ell$ is the jump in the generalized-force density.
For a strong first-order transition, intrabranch excitations remain
gapped.  In the low-temperature coexistence window,
\begin{equation}
    Z_\alpha(n\beta)
    =d_\alpha\e^{-n\beta E_\alpha(g,L)}
    \left[1+\Order{\e^{-n\beta\Delta_\alpha}}\right],
    \label{eq:restrictedBranchZ}
\end{equation}
where $d_\alpha$ is a finite multiplicity and $\Delta_\alpha>0$ is an
intrabranch gap.  Neglecting exponentially small interface and
intrabranch corrections, the partition function can be approximated
as a sum of contributions from the two coexisting
branches~\cite{binder1984finite}:
\begin{equation}
    Z(n\beta,g,L)
    \simeq d_1\e^{-n\beta E_1(g,L)}
    +d_2\e^{-n\beta E_2(g,L)} .
    \label{eq:FOtwoBranchZ}
\end{equation}

Define the energy difference between the two branches as
\begin{equation}
    \Delta E(g,L)=E_2(g,L)-E_1(g,L),
    \qquad
    y=\beta\Delta E(g,L) .
    \label{eq:FOy}
\end{equation}
Factoring out branch 1, the partition function becomes
\begin{equation}
    Z(n\beta)=\e^{-n\beta E_1}
    \left[d_1+d_2\e^{-ny}\right] .
    \label{eq:FOfactoredZ}
\end{equation}
The common branch-1 exponential cancels from the RTE, giving the
two-branch expression
\begin{equation}
    S^{(n)}(y)
    =\frac{
      \ln\left(d_1+d_2\e^{-ny}\right)
      -n\ln\left(d_1+d_2\e^{-y}\right)
    }{1-n} .
    \label{eq:FOSGeneralmain}
\end{equation}
At the branch crossing ($y=0$),
\begin{equation}
    S^{(n)}(0)=\ln(d_1+d_2) .
    \label{eq:FOSatCrossing}
\end{equation}
Thus the RTE itself does not vanish at a first-order transition;
in the nondegenerate two-branch case it equals $\ln2$ at coexistence.

\begin{figure*}[!htp]
\centering
\includegraphics[width=\textwidth]{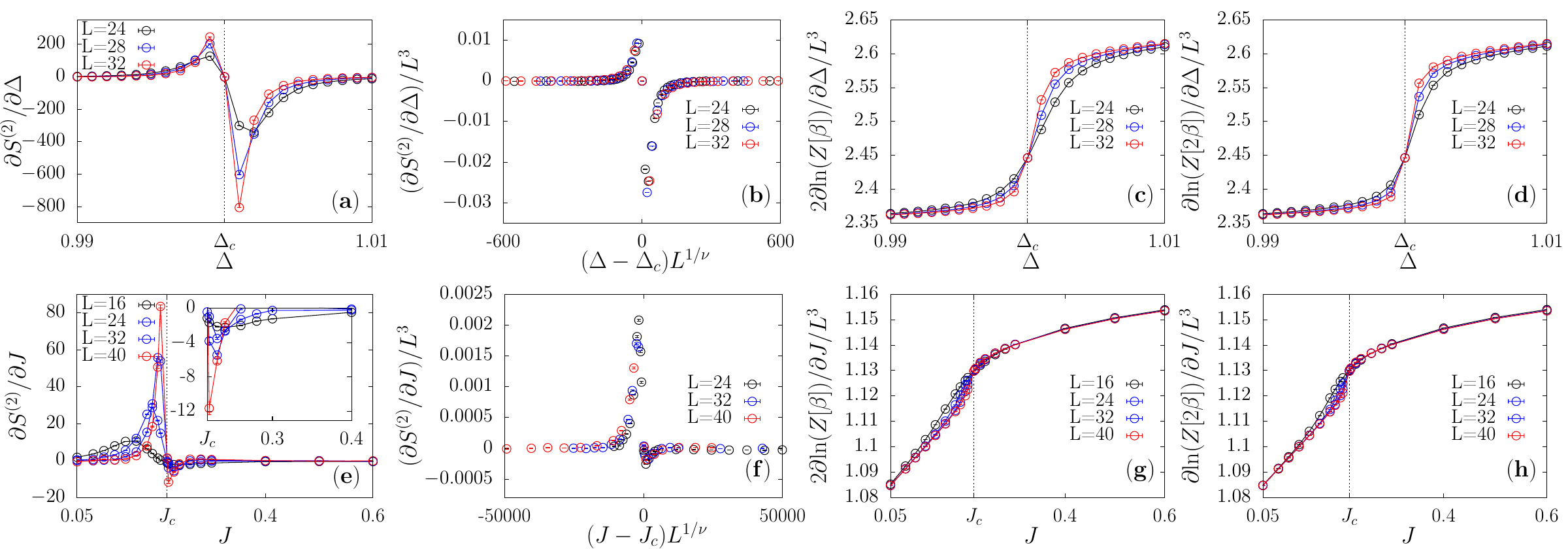}
\caption{(a) Derivative of the  RTE, and (b) corresponding data collapse analysis near the phase transition point of the 2D anisotropic Heisenberg model with anisotropy parameter $\Delta$ [ see Eq. (\ref{anisotropicHeisenberg }) ]. (c) and (d) show the data for the derivative of the partition function at temperatures $\beta$ and $2\beta$ near the transition. Here, $J_1 = J_2 = 1$, and the anisotropy $\Delta$ is taken as the tuning parameter. (e–h) Results for the checkerboard $J$-$Q$  model, using $Q=1$ and tuning $J$. }
\label{fig:firstorder}
\end{figure*}

The probability of branch 2 in the $n\beta$ ensemble  is described by a logistic function
\begin{equation}
    p_n(y)
    =\frac{d_2\e^{-ny}}{d_1+d_2\e^{-ny}},
    \qquad n>0 .
    \label{eq:FOoccupation}
\end{equation}
Differentiating Eq.~\eqref{eq:FOSGeneralmain} and using
$\partial_g y = \beta\partial_g\Delta E(g,L)$, we obtain the DRTE
for general $n$:
\begin{equation}
    \partial_gS^{(n)}
    =\frac{n}{n-1}\,
    \partial_gy\,[p_n(y)-p_1(y)] .
    \label{eq:FODRTEgeneral}
\end{equation}
For $n=2$, this reduces to the compact expression
\begin{equation}
    \partial_gS^{(2)}
    =2\beta\partial_g\Delta E(g,L)
    [p_2(y)-p_1(y)] .
    \label{eq:FODRTE2exact}
\end{equation}
Using Eq.~\eqref{eq:FObranchEnergies1} and retaining the leading
volume-law term, \textbf{the most important result in this section---the key scaling form of the first-order transition---is obtained:}
\begin{equation}
    \partial_gS^{(2)}
    =2\beta L^d\Delta e'(g)
    [p_2(y)-p_1(y)]+\cdots,
    \qquad \Delta e=e_2-e_1 ,
    \label{eq:FODRTE2leadingmain}
\end{equation}
which immediately explains
\textbf{3 characteristic features of the DRTE at a first-order
transition}:

(1) Since $\Delta e'(\gc)=\ell\neq0$, the difference
$p_2(y)-p_1(y)$ changes sign across $\gc$ (near $\gc$,
$y\approx\beta L^d\ell(g-\gc)$, crossing zero linearly), resulting
in a \textbf{robust double-peak structure} with one positive and one
negative peak located on opposite sides of the transition.  This
structure reflects the rapid transfer of thermodynamic weight
between the two competing branches and is a direct consequence of
phase coexistence.

(2) At the first-order transition point ($y=0$), since $    p_n(0)=p_1(0)=\frac{d_2}{d_1+d_2}$, thus
\begin{equation}
    \partial_gS^{(2)}\big|_{\Delta e=0}=0 .
    \label{eq:FOzeroCrossing}
\end{equation}
\textbf{The DRTE crosses through 0 at the first-order transition point.} For unequal multiplicities, the occupation at coexistence is not
generally $1/2$; the zero follows because the $\beta$ and $n\beta$
ensembles have the same branch occupation at the exact branch-energy
crossing.  This represents another fundamental distinction from
continuous phase transitions, where the DRTE scales as $L^{1/\nu}$
without a generic zero crossing.

(3) Because $|p_2(y)-p_1(y)|$ is an $O(1)$ constant and
$\Delta e'(g)$ is also of constant order, the peak amplitude of the
DRTE scales as
\begin{equation}
    \left|\partial_gS^{(2)}\right|
    \sim \beta L^d \sim L^{d+z_\beta}.
\end{equation}
Here $z_\beta$ denotes the power with which the inverse temperature
is increased along the thermal trajectory $\beta = cL^{z_\beta}$
($c>0$).  At a first-order transition there need not be a universal
dynamical exponent; $z_\beta$ is a chosen scaling parameter.

The thermal rounding window is $|y|=O(1)$, which gives
\begin{equation}
    \delta g\sim L^{-(d+z_\beta)},
    \qquad
    \left|\partial_gS^{(2)}\right|_{\rm peak}
    \sim L^{d+z_\beta} .
    \label{eq:FOwidthHeight}
\end{equation}
\textbf{Consequently, the DRTE satisfies the finite-size scaling ansatz}
\begin{equation}
    \partial_gS^{(2)}
    =L^{d+z_\beta}
    f \!\left(\Delta g L^{d+z_\beta};d_2/d_1\right)
    +\cdots .
    \label{eq:FOFSSmain}
\end{equation}
To maintain consistency with the formalism of continuous phase transitions, we can define $1/\nu = d + z_\beta$ for this discontinuity scaling. However, this definition does not imply a diverging correlation length or the existence of a universal first-order dynamical exponent. The divergence of the DRTE does not arise from a critical universality; rather, it stems from the system's intrinsic volume-law scaling and the fact that the free energies of the coexisting phases become unequal once the system moves away from the transition point (see Appendix~\ref{sec:firstorder} for
the more detailed derivation).

\subsection{Numerical demonstration}
Our numerical simulations confirm our theoretical derivation. Specifically, we study the 2D anisotropic Heisenberg model and the 2D checkerboard $J$-$Q$ (CBJQ) model with first-order transitions~\cite{ZhaoBowen2019}. Let us begin by examining the 2D anisotropic Heisenberg model defined in  Eq. (\ref{anisotropicHeisenberg }), where the antiferromagnetic  couplings are uniform, $J_1 = J_2 = 1$, and the anisotropy parameter $\Delta$ serves as the tuning variable. At $\Delta = 1$, the system exhibits full spin-rotation symmetry, corresponding to O(3), which is spontaneously broken by the emergence of AF order in the ground state. When $\Delta$ deviates from unity, the symmetry is explicitly lowered to in-plane spin rotation symmetry and spin-flip symmetry. For $0 < \Delta < 1$, the ground state develops an easy-plane AF order, spontaneously breaking the continuous $\mathrm{SO}(2)$ symmetry. In contrast, for $\Delta > 1$, the system favors an easy-axis AF order, leading to the breaking of the discrete $\mathbb{Z}_2$ spin-inversion symmetry. Precisely at $\Delta = 1$, both types of order coexist, and each vanishes discontinuously across the transition. This abrupt change in the order parameters signals a first-order quantum phase transition, accompanied by the spontaneous breaking of an enhanced O(3) symmetry~\cite{ZhaoBowen2019}.

The DRTE across the transition at $\Delta_c = 1$ is shown in Fig.~\ref{fig:firstorder}(a). A clear discontinuity is observed near the phase transition point, with the DRTE diverging toward opposite signs on either side of the transition in the thermodynamic limit ($L \to \infty$). For comparison, we also compute the derivative of the logarithm of the partition function, as displayed in Figs.~\ref{fig:firstorder}(c) and (d). These curves exhibit a significantly smoother behavior across the transition compared to the DRTE. Moreover, the DRTE curves for different system sizes are seen to intersect precisely at the critical point $\Delta_c = 1$, and the value at the crossing point is zero—consistent with the expected vanishing of the DRTE at a first-order transition. Finally, as shown in the data collapse in Fig.~\ref{fig:firstorder}(b), the DRTE obeys the scaling form given by Eq.~(\ref{eq:FOFSSmain}). These results provide strong numerical confirmation of our theoretical predictions.

The occurrence of a symmetry-enhanced first-order transition is not limited to systems with exact symmetries. An emergent enhanced symmetry can arise at long wavelengths provided that all explicit symmetry-breaking perturbations are irrelevant under the renormalization group (RG) and thus decay along the RG flow. In this scenario, we study the 2D checkerboard $J$-$Q$ model (see Appendix \ref{sec:appmodel} for model details), which exhibits a spontaneously broken emergent O(4) symmetry at the first-order transition point $J_c/Q = 0.217(1)$, where the system transitions from an O(3)-symmetry-breaking antiferromagnetic order to a $\mathbb{Z}_2$-symmetry-breaking plaquette-singlet solid order \cite{ZhaoBowen2019}. The DRTE across this transition is shown in Fig.~\ref{fig:firstorder}(e). A clear discontinuity is observed at the critical point, with the DRTE diverging to opposite signs on either side of the transition in the thermodynamic limit ($L \to \infty$). The DRTE equals zero at the transition point. Furthermore, the DRTE data collapse successfully onto a single curve when rescaled according to Eq.~(\ref{eq:FOFSSmain}) (Fig.~\ref{fig:firstorder}(f)), consistently confirming the validity of our theoretical predictions. Similar to the anisotropic Heisenberg model, the derivative of the logarithm of the partition function (shown in Figs.~\ref{fig:firstorder}(g) and (h)) exhibits a much smoother behavior across the transition compared to the DRTE.  

Combining theoretical analysis with numerical simulations, we demonstrate that the DRTE is not only effective for characterizing continuous phase transitions, but also serves as a powerful probe for first-order phase transitions.

\section{weakly first-order transitions}  
\subsection{Finite-size scaling form}
Having established the universal behaviors at continuous and strong
first-order transitions, we now turn to weakly first-order transitions,
which pose significant challenges for conventional local observables.
The fundamental difficulty is that the correlation length at the
transition, although finite, is much larger than the accessible system
sizes ($L \ll \xi$).  In this regime, the renormalization-group flow
can remain close to a nearby critical fixed point for many decades in
length scale, and observables display pseudocritical scaling with
effective exponents $\nu$ and $z$ controlled by that nearby critical
point---even though the thermodynamic transition is first
order\cite{gorbenko2018walking,gorbenko2018walking2}

In the standard first-order DRTE, quantum fluctuations were neglected
because intrabranch excitations are gapped.  For a weakly first-order
transition, however, the proximity of a critical point introduces
strong quantum fluctuations that must be properly accounted for.
The finite-size spectrum therefore contains two ingredients
simultaneously: two competing macroscopic branches and low-energy
tower corrections inherited from the nearby scaling regime.
Let $\alpha=1,2$ label the two branches and $m$ label the low-energy
tower attached to branch $\alpha$.  In the critical region of the nearby critical point, the
branch-resolved energies take the form
\begin{equation}
    E_{\alpha m}(g,L)
    =L^d e_\alpha(g)+L^{-z}\varepsilon_{\alpha m}(x)+\cdots,
    \quad x=(g-g_c)L^{1/\nu}.
    \label{eq:WFOspectrum}
\end{equation}
where $\varepsilon_{\alpha m}(x)$ is a dimensionless scaling function
that encodes the low-energy finite-size level $m$ of branch $\alpha$
inherited from the nearby critical regime. At inverse temperature $\beta=cL^{z}$, the restricted partition
function of each branch is
\begin{equation}
\begin{split}
    Z_\alpha(n\beta)
    &=\e^{-n\beta L^d e_\alpha(g)}Q_\alpha(nc,x),\\[2pt]
    Q_\alpha(nc,x)
    &=\sum_m\e^{-nc\varepsilon_{\alpha m}(x)}+\cdots,
\end{split}
    \label{eq:WFOrestrictedZ}
\end{equation}
where the tower factor $Q_\alpha(nc,x)$ is $O(1)$ along the finite-size
trajectory.  The full two-branch partition function is therefore
\begin{equation}
    Z(n\beta)
    \simeq
    \e^{-n\beta L^d e_1(g)}Q_1(nc,x)
    +\e^{-n\beta L^d e_2(g)}Q_2(nc,x) .
    \label{eq:WFOtwoTowerZ}
\end{equation}

\begin{figure*}[htp]
\centering
\includegraphics[width=\textwidth]{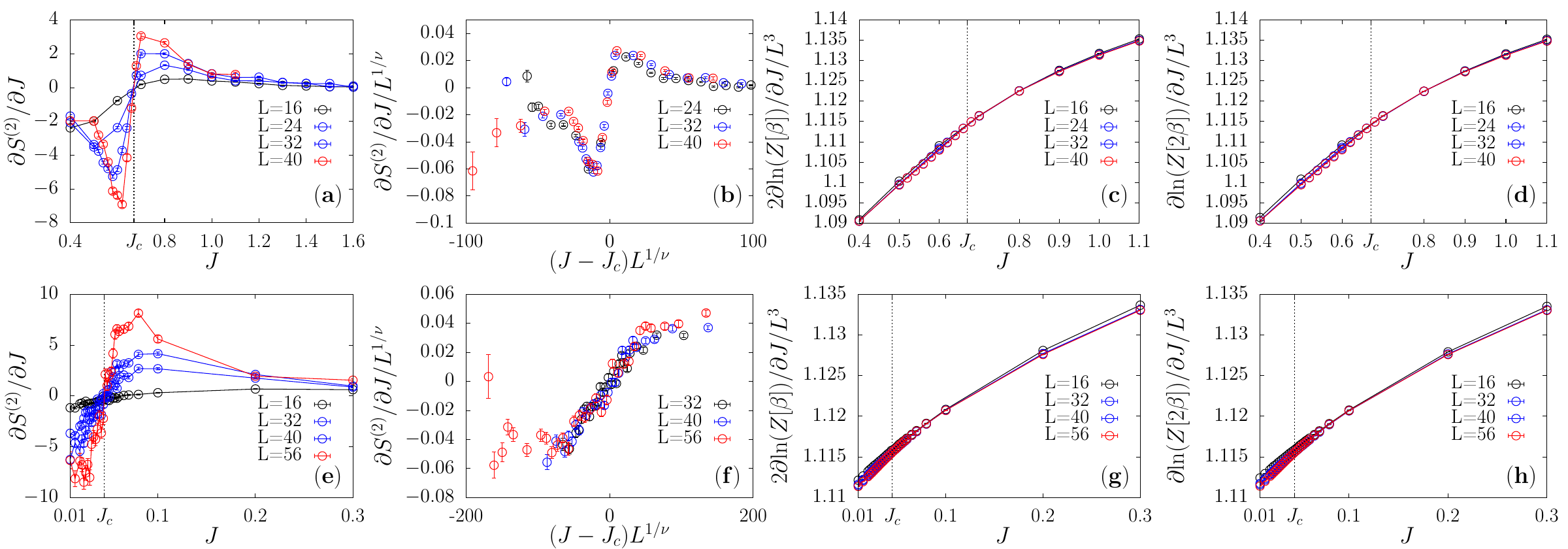}
\caption{We set $Q=1$ and use $J$ as the tuning parameter. (a) The derivative of the  RTE: DRTE $\partial S^{(2)}/\partial J$.  (b) Corresponding data collapse analysis of (a).
 (c) and (d) show the data for the derivative of the partition function at temperatures $\beta$ and $2\beta$ near the transition of the $J-Q_3$ model. (e--h) Data for the $J-Q_2$ model.  The raw values of the partition function derivative vary significantly across system sizes (see Appendix \ref{sec:appmodel}), making direct comparison difficult. To resolve this, we normalize the data by $ L^3 $, corresponding to $ \beta L^2 = L^3 $.}
\label{fig:wft}
\end{figure*}

Unlike in the strong first-order limit, the tower factors \(Q_\alpha\)
cannot in general be treated as constants, because each competing branch
retains low-energy spectral weight inherited from the nearby critical fixed
point. Consequently, the DRTE combines the branch-occupation response
characteristic of a first-order transition with the finite-size response of
the pseudocritical tower. After factoring out the common extensive reference
energy, and provided that the residual branch-independent tower contribution
is negligible or has been subtracted, \textbf{the leading DRTE of a weakly first-order transition can be written in the compact form}
\begin{equation}
    \frac{\partial S^{(2)}}{\partial g}
    = \Bigl[
       2cL^{d+z_{\beta}}\Delta e'(g)
       + \widetilde S'(x)L^{1/\nu}
     \Bigr]
     [p_2(y)-p_1(y)]
     +\cdots,
    \label{eq:WFOmainTextForm1}
\end{equation}
where the mark $'$ means the derivative versus $g$. A detailed derivation and the conditions underlying this expression are given
in Appendix~\ref{sec:weaklyfirstorder}.

\textbf{Equation~\eqref{eq:WFOmainTextForm1} is the central result for weakly
first-order transitions.  It predicts three characteristic features}:

(1) The \textbf{double-peak structure} of a first-order transition
is preserved even when the transition is weak, since the occupation difference $p_2(y)-p_1(y)$
remains.

(2) The \textbf{DRTE is 0 at the finite-size coexistence point},
where $p_2(0)=p_1(0)=d_2/(d_1+d_2)$.  This zero crossing is the
same thermodynamic identity as in the strong first-order case.

(3) For accessible system sizes $L \ll \xi$, the coefficient
$\Delta e'(g)$ is parametrically small because the two branches are
nearly degenerate near the nearby critical fixed point.  Consequently,
the DRTE does not follow the $L^{d+z_{\beta}}$ scaling characteristic of a
strong first-order transition; instead, it exhibits $L^{1/\nu}$
pseudocritical scaling inherited from the nearby critical point.
When $L \gg \xi$, however, the finite correlation length cuts off the
critical fluctuations of nearby critical fixed point, and the DRTE crosses over to the $L^{d+z}$
scaling of a standard first-order transition.

In the following, we present numerical data that strongly support the
analysis above.

\subsection{Finite-size advantage and range of validity}

We now clarify why the DRTE can reveal the first-order component of a
weakly first-order transition at relatively small system sizes, and the
finite-size regime in which this diagnosis is controlled.  The favorable
finite-size behavior of the DRTE has two complementary physical origins.

First, the Rényi construction suppresses a large smooth contribution
to the partition function before any finite-size fitting is performed.
Schematically, the logarithm of the partition function can be written as
\begin{equation}
\ln Z(\beta,g,L)
=
\beta E_{\rm reg}(g,L)
+
\mathcal U(\beta,g,L)
+
\delta\mathcal A(\beta,g,L),
\label{eq:WFObackground}
\end{equation}
where $E_{\rm reg}(g,L)$ denotes the dominant smooth contribution that
is extensive in the imaginary-time direction, $\mathcal U$ contains the
nontrivial long-wavelength finite-size contribution, and
$\delta\mathcal A$ represents remaining subleading corrections.  The
regular contribution may contain the spatial expansion
\begin{equation}
E_{\rm reg}(g,L)
=
a_0(g)L^d
+
a_1(g)L^{d-1}
+
a_2(g)L^{d-2}
+\cdots .
\end{equation}
In the Rényi combination,
\begin{equation}
S^{(n)}
=
\frac{1}{1-n}
\left[
\ln Z(n\beta)-n\ln Z(\beta)
\right],
\end{equation}
the contribution that is linear in the imaginary-time extent cancels
identically.  Therefore, the large smooth background common to the two
thermal partition functions is strongly suppressed, making the
remaining long-wavelength finite-size structure more directly visible.
This cancellation does not imply that all finite-size corrections
disappear: corrections from irrelevant operators, branch-dependent
low-energy spectra, and other subleading terms can remain.

Second, once two competing thermodynamic branches contribute
appreciably to the finite-size partition function, the DRTE directly
probes the redistribution of their statistical weights between the
$\beta$ and $n\beta$ ensembles.  As shown by the two-branch description
above (Equation~\eqref{eq:WFOmainTextForm1}).
The occupation difference $p_2(y)-p_1(y)$ changes sign across the
branch crossing and vanishes at the finite-size coexistence point.
This directly generates the zero crossing and opposite-sign
double-peak structure.  For a weakly first-order transition, these
first-order occupation signatures can coexist with the pseudocritical
$L^{1/\nu}$ scaling inherited from the nearby critical fixed point.
The combination of background suppression and direct sensitivity to
branch redistribution explains why the first-order component can become
visible in the DRTE even while other observables remain dominated by
critical-like finite-size behavior.

These mechanisms follow from the structure of the Rényi construction
and are not specific to the particular models studied here.  However,
they do not imply that  arbitrarily
small systems are sufficient.  The numerical size at which the
characteristic DRTE structure becomes clearly resolvable remains
model dependent.

Indeed, the scaling theory developed above is a low-energy effective
description and therefore requires a separation between the
long-wavelength degrees of freedom controlling the transition and
microscopic short-distance excitations.  Along the finite-size
trajectory
\begin{equation}
\beta=cL^z,
\end{equation}
a low-energy excitation with
\begin{equation}
\Delta_{\rm IR}(L)\sim L^{-z}
\end{equation}
satisfies
\begin{equation}
\beta\Delta_{\rm IR}=O(1),
\end{equation}
and therefore remains active in the finite-size partition function.
By contrast, a microscopic excitation has a finite energy scale
$\Delta_{\rm UV}$ and contributes with a Boltzmann factor
\begin{equation}
e^{-\beta\Delta_{\rm UV}}
=
e^{-cL^z\Delta_{\rm UV}}.
\end{equation}
For sufficiently large $L$,
\begin{equation}
\beta\Delta_{\rm UV}
=
cL^z\Delta_{\rm UV}
\gg1,
\end{equation}
so that these microscopic excitations are strongly suppressed.
Equivalently, in space one requires $L\gg a$, where $a$ is the
microscopic lattice scale.

We denote by $L_{\min}$ the model-dependent crossover scale beyond
which this low-energy separation of scales becomes effective.
Schematically, its physical meaning is set by the conditions
\begin{equation}
cL^z\Delta_{\rm UV}\gg1,
\qquad
L/a\gg1.
\label{eq:LminCondition}
\end{equation}
The scale $L_{\min}$ is not universal: it depends on the ultraviolet
energy scale, lattice cutoff, irrelevant couplings, and other
short-distance details.  It is also physically distinct from the
correlation length $\xi$.  The former characterizes the crossover from
microscopic physics to the low-energy effective regime, whereas the
latter characterizes the eventual departure from pseudocritical
behavior toward the asymptotic first-order regime.

For a weakly first-order transition with an anomalously large
correlation length, these two scales naturally define an intermediate
window
\begin{equation}
L_{\min} \ll L \ll \xi.
\label{eq:WFOwindow}
\end{equation}
The lower inequality ensures that microscopic high-energy contributions
are sufficiently suppressed for the low-energy scaling theory to apply,
whereas the upper inequality keeps the system in the pseudocritical
regime controlled by the nearby critical fixed point.  It is precisely
in this window that the DRTE can simultaneously display pseudocritical
$L^{1/\nu}$ scaling and the zero-crossing/double-peak structure
associated with first-order branch competition.

If $L\lesssim L_{\min}$, microscopic and low-energy physics are still
strongly mixed.  Short-range lattice-dependent contributions can
distort the scaling function, shift or round the zero crossing, and
obscure the double-peak structure.  A weakly first-order transition may
therefore resemble a continuous transition at such small sizes.
Importantly, the absence of a first-order DRTE signature in this regime
should be regarded as inconclusive rather than as evidence for a
continuous transition.  This limitation is not specific to the DRTE:
in conventional finite-size analyses of order parameters, correlation
functions, Binder cumulants, and susceptibilities, the smallest system
sizes are likewise excluded when microscopic corrections prevent a
stable scaling description.

Thus, the finite-size advantage of the DRTE is a general physical
mechanism, whereas the numerical crossover scale at which this
advantage becomes visible is model dependent.  In the following, we
show that the smooth
partition-function responses remain nearly featureless while the DRTE
already resolves the characteristic first-order branch structure.

\subsection{Numerical demonstration}

The phase transition between AF N\'eel order and VBS phase in 2D quantum magnets can appear (approximately) continuous, which is thought as the deconfined quantum criticality (DQC)~\cite{senthilDeconfined2004, senthilQuantum2004}. $J-Q_2$ and $J-Q_3$ models are prototypical examples that realize such transitions in 2D quantum spin systems~\cite{sandvik2007evidence,lou2009antiferromagnetic}. These models are free of the sign problem, making them ideal for QMC simulations.  Extensive QMC studies have shown that the phase transitions of the $J-Q$ models on the 2D lattice are continuous and exhibit emergent $SO(5)$ symmetry, which is a hallmark of DQC~\cite{sandvik2007evidence,lou2009antiferromagnetic,Melko2008Scaling,Kaul2011Quantum,shao2016quantum,Nahum2015Deconfined}. However, the critical exponents observed in these models do not satisfy the conformal bootstrap bounds expected for SO(5) critical point~\cite{Nakayama2016Necessary}. Recent works have increasingly suggested that this DQC scenario may instead correspond to a weakly first order transition ~\cite{JRZhao2021,Zhao2020Multicritical,Lu2021Self,Kuklov2008Deconfined,Chen2013Deconfined,deng2024diagnosing,zhou2024mathrmso5,Jun2024SO(5),Yuan2023,Menghan2024SciPost,Menghan2025}.
Next, we will show that the DRTE behaviors provide a clear and direct signature that the transitions in the $ J- Q_2 $ and $ J - Q_3 $ models  ( see Appendix \ref{sec:appmodel} for models' details) are indeed weakly first order in finite sizes (up to $ L = 56 $). The observed DRTE signals fully align with the three key features predicted in the previous subsection, offering compelling evidence against a genuine continuous transition.

The transition points of the 2D square lattice 
$J-Q_2$ and $J-Q_3$ models have been numerically determined as $J_c/Q= 0.04502$ and $J_c/Q=0.67046$, respectively~\cite{sandvik2007evidence,lou2009antiferromagnetic,YCWang2021DQCdisorder,Jun2024SO(5)}. The DRTE results for both models are shown in Figs. \ref{fig:wft} (a) and (e),  along with the derivatives of the logarithmic partition function in panels (c,d) and (g,h).  The derivative of the logarithmic partition function appears smooth across the transition, even for large system sizes, as seen in Figs.~\ref{fig:wft}(c,d) for $ J- Q_3 $ and (g,h) for $ J-Q_2 $ model. This shows that the almost smooth derivative of the logarithmic partition function cannot directly reveal the weakly first-order nature, as discussed above, since the logarithmic partition function contains large analytic background contributions.

In contrast, the DRTE reveals clear discontinuities with a double-peak character near the transition points (see Figs. \ref{fig:wft} (a) and (e)), as the feature 1) we predicted in the above subsection. Moreover, DRTE curves for different system sizes intersect at zero at the transition point (as the feature 2) we predicted above), unlike in continuous phase transitions. These two features are fully consistent with our previous analysis of the DRTE at weakly and standard first order transitions. Our results demonstrate that the DRTE can unambiguously detect the first-order nature of weakly first-order phase transitions in relatively small systems (up to $L = 56$), even in the $J$-$Q_2$ model with much stronger finite-size effects. In fact, its nature, whether continuous or first-order, remains under debate so far~\cite{Jun2024SO(5)}.

We now verify our third theoretical prediction for weakly first-order transitions. We first attempted to scale the DRTE data according to Eq.~(\ref{eq:FOFSSmain}), assuming standard first-order scaling. However, we found that the data from different system sizes do not collapse onto a single curve, indicating the inadequacy of this scaling form. We then turned to the modified scaling ansatz for weaklyly first-order transitions, given by Eq.~(\ref{eq:WFOmainTextForm1}), using the initially estimated deconfined quantum critical exponent $\nu = 0.78$~\cite{sandvik2007evidence}. Remarkably, for both the $J$–$Q_2$ and $J$–$Q_3$ models, excellent data collapse is achieved when this form is applied (see Figs.~\ref{fig:wft}(b) and~\ref{fig:wft}(f)). This scaling behavior is fully consistent with the theoretical expectations discussed in the previous subsection—particularly feature 3)—and provides strong evidence that the transitions exhibit the characteristic physics of a weaklyly first-order process, dominated by proximity to a nearby critical point.

It is worth noting that, in current studies based on local order
parameters, the observed behavior within accessible system sizes is
predominantly governed by signatures of continuous phase transitions,
making it difficult to unambiguously identify the underlying first-order
nature.  As a concrete comparison, the Binder cumulant for the
$J$-$Q_2$ model shows no negative values even up to $L=256$, and the
spin and dimer correlation functions measured up to $L=384$ do not
display a clear tendency to converge to finite constants (see
Appendix~\ref{Binderandcorrelation} for details); the computational cost
of reaching $L=384$ is roughly 300 times that of $L=56$.  In stark
contrast, our DRTE already provides a definitive, qualitative signature
of the weakly first-order transition at $L=56$.  This provides a clear and
compelling demonstration of the weakly first-order character of the
transition, marking, to our knowledge, the first such identification
achieved at modest system sizes.

As an additional test, we apply the same DRTE diagnostic to the square-lattice SU(8) $J_1$--$J_2$ model, with the detailed results presented in Appendix~\ref{app:SU(8)}. Recent smooth-boundary entanglement-entropy scaling was interpreted as being compatible with a continuous deconfined quantum critical point in this model~\cite{song2025evolution}. In contrast, our DRTE data exhibit a clear sign-changing structure with a zero crossing and opposite-sign extrema near the transition, while the same data simultaneously admit a pseudocritical scaling collapse with an effective exponent $\nu=0.67(4)$. Within the framework developed above, this coexistence of critical-like finite-size scaling and first-order branch-competition signatures is characteristic of a weakly first-order transition. The SU(8) $J_1$--$J_2$ model therefore provides an additional nontrivial application of the DRTE, showing that it can reveal a first-order component in a case where other finite-size probes have been interpreted as being consistent with continuous critical behavior.

\section{Conclusion and discussions}
We have established a unified  scaling theory for the R\'enyi thermal entropy (RTE) and its derivative (DRTE) across all three fundamental types of phase transitions: continuous (second-order), strong first-order, and weakly first-order. Our framework is predicated on the key insight that the DRTE naturally isolates the universal contribution of the free energy by canceling dominant analytic background contributions. Crucially, we have derived distinct and diagnostic scaling behaviors for the DRTE in each regime:

1. \textbf{Continuous Phase Transitions:} The DRTE scales as
   \begin{equation}
   \frac{\partial S^{(2)}}{\partial g} = L^{1/\nu} \, \tilde{S}'(\Delta g L^{1/\nu}),
   \end{equation}
   where $\nu$ is the correlation length exponent. This allows for precise determination of the critical point $g_c$ and exponent $\nu$ via data collapse, as demonstrated for (2+1)D O($N$) universality classes.

2. \textbf{Standard First-Order Transitions:} The DRTE exhibits a characteristic discontinuity with a double-peak structure and follows the volume-law scaling
  \begin{equation}
    \partial_gS^{(2)}
    =L^{d+z_\beta}
    f \!\left(\Delta g L^{d+z_\beta};d_2/d_1\right)
    +\cdots .
\end{equation}
   where $d$ is the spatial dimension and $z_\beta$ is a chosen scaling parameter (at a first-order transition there need not be a universal
dynamical exponent). The zero crossing at $g_c$ and the sign change of the peaks provide an unambiguous signature of phase coexistence.

3. \textbf{Weakly First-Order Transitions:} The DRTE obeys a hybrid scaling form
   \begin{equation}
    \frac{\partial S^{(2)}}{\partial g}
    = \Bigl[
       2cL^{d+z_{\beta}}\Delta e'(g)
       + \widetilde S'(x)L^{1/\nu}
     \Bigr]
     [p_2(y)-p_1(y)]
     +\cdots,
\end{equation}
   which encapsulates the competition between the underlying first-order mechanism (first term) and the influence of a nearby critical fixed point (second term). In accessible finite sizes ($L \ll \xi$), the critical-like $L^{1/\nu}$ scaling dominates, explaining the observed data collapse with continuous exponents. Simultaneously, the persistent double-peak structure and zero crossing betray the first-order nature, as conclusively shown for the debated $J$-$Q$ models.

   Our numerical results establish this framework at two complementary levels. First, the well-established O(1), O(2), and O(3) quantum critical points and the strong first-order anisotropic Heisenberg and checkerboard $J$-$Q$ transitions provide controlled benchmarks for the continuous- and first-order scaling theories.  Second, the $J$-$Q_2$, $J$-$Q_3$, and SU(8) $J_1$-$J_2$ models constitute nontrivial applications where the transition character is considerably more subtle. In all three cases, the DRTE reveals the coexistence of pseudocritical scaling and first-order branch-competition signatures. In particular, the result of SU(8) $J_1$-$J_2$ model indicates weakly first-order behavior even though recent smooth-boundary entanglement-entropy scaling was interpreted as being compatible with a continuous transition~\cite{song2025evolution}. These results therefore go beyond benchmark confirmation of the scaling theory and demonstrate its ability to uncover previously unresolved first-order components in candidate deconfined quantum critical systems.

Beyond these specific findings, the RTE/DRTE approach offers a
general and order-parameter-agnostic diagnostic tool.  Its
applicability extends to transitions elusive to conventional local
observables, including potential topological phase transitions.
While validated here with high-precision quantum Monte Carlo, our
method is readily adaptable to other numerical techniques.  As a
concrete demonstration, we have implemented our scaling framework
within the thermal tensor network method for the one-dimensional
transverse-field Ising model, successfully extracting the critical
point and exponent $\nu$ in agreement with exact results (see
Appendix~\ref{Tensor} for details)~\cite{Chen2018Exponential,Li2023Tangent}.
This work thus provides a new paradigm for characterizing quantum
criticality, resolving the long-standing challenge of distinguishing
weakly first-order from continuous transitions in finite-size
simulations.

\begin{acknowledgements}
Zhe Wang and Yanzhang Zhu contributed equally to this work. We thank the helpful discussions with Cheng-Xiang Ding, Yan-Cheng Wang, Menghan Song, Zi Yang Meng, Chong Wang and Wenan Guo. 
This project is supported by the Scientific Research Project (No. WU2025B011), Feng-Ying Career Development Chair Fund and the Start-up Funding of Westlake University, and the Natural Science Foundation of China (Grant No. 12674188), and Quantum Meta Key (QuaMeK) computing. The authors thank the IT service office and the high-performance computing center of Westlake University.
\end{acknowledgements}

\clearpage
\appendix
\setcounter{equation}{0}
\setcounter{figure}{0}
\renewcommand{\theequation}{S\arabic{equation}}
\renewcommand{\thefigure}{S\arabic{figure}}
\setcounter{page}{1}
\begin{widetext}
	
\section{Method} 
\label{sec:appm}
The $n$th order R\'enyi entropy is defined as $S^{(n)} =\frac{1}{1-n}\ln \mathrm{Tr}(\rho^{n})$ (or $\frac{1}{1-n}\ln \mathrm{Tr}(\rho_A^{n})$), where $\rho=e^{-\beta H}/Z$  and $Z=\mathrm{Tr}e^{-\beta H}$ (with $H$ being the Hamiltonian ).  Here, $\rho_A=\mathrm{Tr}_B \rho$ represents the reduced density matrix of subsystem $A$ coupled with an environment $B$. The R\'enyi entanglement entropy is defined using $\rho_A$, while the   R\'enyi thermal entropy (RTE) is based on  $\rho$. In this work, we focus on RTE. From the above definitions, it is evident that the R\'enyi entropy can be expressed as the ratio of partition functions at two different temperatures: $S^{(n)} =\frac{1}{1-n}\ln\frac{Z(n\beta)}{Z(\beta)^n} $ (for simplicity, we consider $S^{(2)}$ in this work).

The partition functions can be calculated using the bipartite reweight-annealing algorithm~\cite{wang2024ee,ding2024reweightannealing,dai2024residual,neal2001annealed} in conjunction with stochastic series expansion (SSE)  QMC simulations~\cite{Sandvik1999,sandvik2010computational,sandvik2019stochastic,Syljuaasen2002,yan2020improved,yan2019sweeping}.  The key idea is to transform the problem of solving the ratio of partition functions with different parameter into the problem of reweighting:
\begin{equation}\label{eq:ratio}
\frac{Z(J')}{Z(J)} = \bigg\langle \frac{W(J')}{W(J)} \bigg\rangle  _{Z(\beta,J) or Z(n\beta,J)}
\end{equation}
where the $\langle ...\rangle_{Z(\beta,J) or Z(n\beta,J)}$ indicates that the QMC samplings is performed under the manifold $Z(\beta,J) $ or $Z(n\beta,J)$ at parameter $J$ (note parameter $J$ is physical parameter here). $W(J')$ and $W(J)$ represent the weights of a same configuration but under different parameters $J'$ and $J$. 

In this frame, we can designed the incremental process along the path composed of real physical parameters. In other words, all intermediates are physical partition functions at corresponding parameters. The continuously incremental trick here are introduce as follows:
\begin{equation}\label{eq:ratio2}
\frac{Z(J')}{Z(J)} = \prod_{i=0}^{N-1} \frac{Z(J_{i+1})}{Z(J_i)}
\end{equation}
where $J_N=J'$ and  $J_0=J$, others $J_i$ locate between the two in sequence. 
In realistic simulation, we are able to gain any ratio ${Z(J')}/{Z(J)}$ in this way. But $Z(n\beta,J')/Z^n(\beta,J')$ can not be obtained. The antidote comes from some
known reference point $Z(n\beta,J)/Z^n(\beta,J)$:   [Assuming that we have calculated the ratios $Z(n\beta,J')/Z(n\beta,J)$ and $Z(\beta,J')/Z(\beta,J)$ using the method described above, the ratio  $Z(n\beta,J')/Z^n(\beta,J')$ can be obtained if  $Z(n\beta,J)/Z^n(\beta,J)$ is known. For a given zero-temperature phase, and assuming the ground state degeneracy is $ d_e $, one can derive that: $S^{(n)} =\frac{1}{1-n}\ln\frac{Z(n\beta)}{Z^n(\beta)} =\ln de$. Therefore, we can use states with known degeneracies — such as dimer product states (with degeneracy 1 in this work) and discrete symmetry-breaking states (with degeneracy 2 in this work) — as reference points at a given parameter $ J $. In this way,  $Z(n\beta,J)/Z^n(\beta,J)$ can be determined by solving:
 $\frac{1}{1-n}\ln\frac{Z(n\beta,J)}{Z^n(\beta,J)} =\ln de$.]

One might be concerned about how to deal with a Hamiltonian without a state with known degeneracy. In this way, we can introduce a temperature axis starting from infinite temperature and annealing down to the target temperature. The degeneracy at infinite temperature is known as $2^N$ (consider an N-spin system with spin-1/2).

The RTE derivative of $J$ is
\begin{equation}
\frac{dS^{(n)}}{dJ}=\frac{1}{1-n}\bigg[\frac{dZ(n\beta)/dJ}{Z(n\beta)}-n\frac{dZ(\beta)/dJ}{Z(\beta)}\bigg]
\label{deri-EE}
\end{equation}
According to $Z(\beta)=\mathrm{Tr}e^{-\beta H}$, we have
\begin{equation}
\frac{dZ}{dJ}=\mathrm{Tr}[-\beta \frac{dH}{dJ} e^{-\beta H}]
\label{deri-z}
\end{equation}
Thus
\begin{equation}
\frac{dZ/dJ}{Z}=-\beta\bigg\langle \frac{dH}{dJ} \bigg\rangle_Z
\label{dzdjz}
\end{equation}
Similarly, we can  obtain
\begin{equation}
\begin{split}
\frac{dZ(n\beta))/dJ}{Z(n\beta))}&=-n\beta\bigg\langle \frac{dH}{dJ} \bigg\rangle_{Z(n\beta)}
\label{dzdjzan}
\end{split}
\end{equation}
Therefore, the RTE derivative can be rewritten as
\begin{equation}
\frac{dS^{(n)}}{dJ}=\frac{1}{1-n}\bigg[-n\beta\bigg\langle \frac{dH}{dJ} \bigg\rangle_{Z(n\beta)}+n\beta\bigg\langle \frac{dH}{dJ} \bigg\rangle_Z \bigg]
\label{deri-EE2}
\end{equation}

The equations above is general and doesn't depend on detailed quantum Monte Carlo (QMC) methods. Then let us discuss how to calculate them in stochastic series expansion (SSE) simulation.
For convenience, we fix the R\'enyi index $n=2$ and choose the dimerized Heisenberg model in main text as the example for explaining technical details.
The Hamiltonian is 
\begin{equation}
H=J_1\sum_{\langle ij \rangle}S_iS_j+J_2\sum_{\langle ij \rangle}S_iS_j
\end{equation}
In the following, we fix $J_2=1$ and leave $J_1$ as the tunable parameter. Note the Hamiltonian is a linear function of $J_1$, that means $dH/dJ_1=H_{J_1}/J_1$ in which $H_{J_1}$ is the $J_1$ term in Hamiltonian. Then the RTE derivative can be simplified as 
\begin{equation}
\frac{dS^{(2)}}{dJ_1}=\bigg[2\beta\bigg\langle \frac{H_{J_1}}{J_1} \bigg\rangle_{Z(2\beta)}-2\beta\bigg\langle \frac{H_{J_1}}{J_1} \bigg\rangle_Z \bigg]
\label{deri-EE3}
\end{equation}
In the SSE frame, it is easy to obtain $\langle H \rangle=\langle -n_{op}/\beta \rangle$~\footnote{How to measure the energy in SSE has been carefully explained in Prof. Sandvik's tutorial \url{http://physics.bu.edu/~sandvik/programs/ssebasic/ssebasic.html}}, where $n_{op}$ is the number of the concerned Hamiltonian operators. Thus the Eq.~(\ref{deri-EE3}) can be further simplified to
\begin{equation}
\frac{dS^{(2)}}{dJ_1}=\bigg[-\bigg\langle \frac{n_{J_1}}{J_1} \bigg\rangle_{Z(2\beta)}+2\bigg\langle \frac{n_{J_1}}{J_1} \bigg\rangle_Z \bigg]
\label{deri-EE4}
\end{equation}
where $n_{J_1}$ means the number of $J_1$ operators including both diagonal and off-diagonal ones. It's worth noting that there is no ``$2$'' anymore in the $\langle ... \rangle_{Z(2\beta)}$ term, because $n_{J_1}$ here contains two replicas' operators which has already been doubled actually.

Note that the RTE is computed using the reweight-annealing QMC method mentioned above, while the DRTE is obtained directly by separately evaluating the two expectation values involved in Eq.~(\ref{deri-EE3}) or Eq.~(\ref{deri-EE4}). This implies that the computational cost of evaluating DRTE is roughly twice that of energy calculations. Typically, we perform on the order of one million Monte Carlo steps when computing the energy, and we have adopted the same number of steps in the present study.

\begin{figure*}[!h]
\centering
\includegraphics[width=0.8\textwidth]{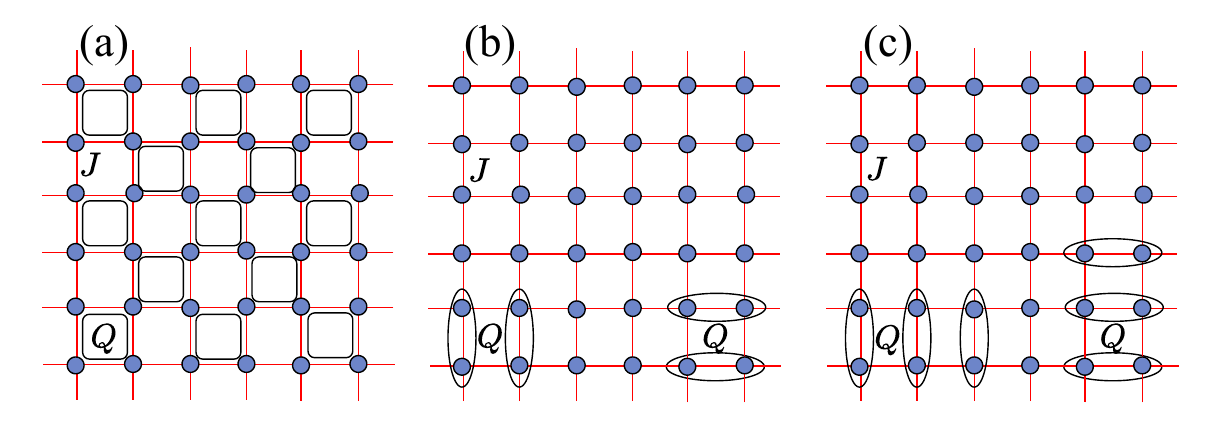}
\caption{$J$-$Q$ models, in which the antiferromagnetic  Heisenberg interaction ($J$-term) acts on all nearest-neighbor bonds. (a) The checkerboard $J$-$Q$ model on a  square lattice with periodic boundary condition in both directions.  The four-spin interaction ($Q$-term) acts on nearest-neighbor bond pairs that form a plaquette. (b) $J-Q_2$ model:  the four-spin $Q$ interaction covers the entire lattice. (c) $J-Q_3$ model:  the six-spin $Q$ interaction covers the entire lattice.}
\label{fig:cbjq}
\end{figure*}
\section{$J-Q$ models} 
\label{sec:appmodel}

In this work, we studied three types of $J$-$Q$ models, namely checkerboard $J$-$Q$, $J$-$Q_2$, and $J$-$Q_3$ models. The Hamiltonian of the checkerboard $J$-$Q$ model on a square lattice is given by~\cite{ZhaoBowen2019}
\begin{equation}
H=-J\sum_{\langle ij \rangle}P_{ij}-Q\sum_{[ijkl]}P_{ij}P_{kl},
\label{eq:cbjq}
\end{equation}
where $P_{ij}=1/4-\vec{S}_{i}\cdot\vec{S}_{j}$  denotes the projection operator onto the singlet state of spins at sites  $i$ and $j$. The first term sums over all nearest-neighbor bond pairs $ \langle ij \rangle $, while the second term runs over plaquettes composed of two parallel bonds $ (ij) $ and $ (kl) $, as illustrated in Fig.~\ref{fig:cbjq}(a). $J$-$Q_2$ and checkerboard $J$-$Q$ model share the same Hamiltonian form, except that in $J$-$Q_2$~\cite{sandvik2007evidence} model, the $Q$ interactions span the entire lattice, as illustrated in Fig.~\ref{fig:cbjq}(b). 
Compared to $J$-$Q_2$ model, the $Q$ interactions in $J$--$Q_3$ model, are composed of the product of three singlet operators~\cite{lou2009antiferromagnetic}:

\begin{equation}
H=-J\sum_{\langle ij \rangle}P_{ij}-Q\sum_{[ijklmn]}P_{ij}P_{kl}P_{mn},
\label{eq:cbjq}
\end{equation}
, as illustrated in Fig.~\ref{fig:cbjq}(c).  The DRTE results for three models are shown in Figs. \ref{fig:wftsm},  along with the  derivatives of the logarithmic partition function in panels (b,c), (e,f) and (h,i).

\begin{figure*}[htb]
\centering
\includegraphics[width=0.8\textwidth]{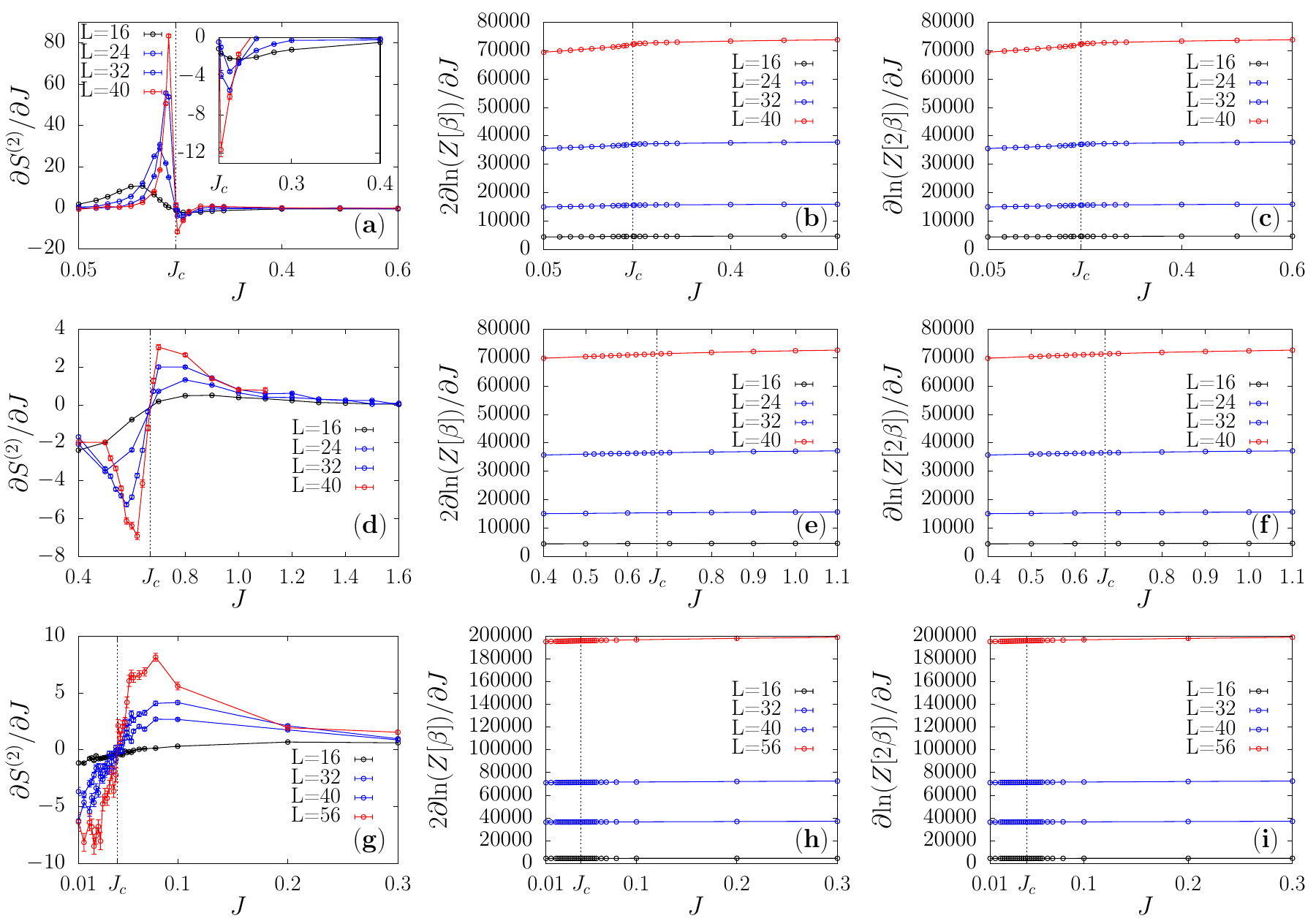}
\caption{We set $Q=1$ and use $J$ as the tuning parameter. (a) The derivative of the  RTE: DRTE $\partial S^{(2)}/\partial J$. An inset shows the data on the right side of the phase transition point. (b) and (c) show the data for the derivative of the partition function at temperatures $\beta$ and $2\beta$ near the transition of the checkerboard $J$-$Q$ model. (d--f) Data for the $J-Q_3$ model. (g--i) Data for the $J-Q_2$ model. }
\label{fig:wftsm}
\end{figure*}

\section{Definitions, thermal protocol, and exact identities}
\label{sec:definitions}

We consider a finite quantum system of linear extent $L$ in $d$ spatial dimensions, governed by a Hamiltonian $H(g)$ that depends smoothly on a tuning parameter $g$. Its partition function and normalized thermal density matrix are
\begin{equation}
    Z(\beta,g,L)=\Tr\,\e^{-\beta H(g)},
    \qquad
    \rho(\beta,g,L)=\frac{\e^{-\beta H(g)}}{Z(\beta,g,L)} .
    \label{eq:defZrho}
\end{equation}
The order-$n$ thermal \Renyi{} entropy is~\cite{renyi1961measures}
\begin{equation}
    S^{(n)}(\beta,g,L)
    =\frac{1}{1-n}\ln\Tr\rho^n
    =\frac{\ln Z(n\beta,g,L)-n\ln Z(\beta,g,L)}{1-n} .
    \label{eq:defRTE}
\end{equation}
For $n=2$,
\begin{equation}
    S^{(2)}=2\ln Z(\beta)-\ln Z(2\beta)
    =-\ln\frac{Z(2\beta)}{Z(\beta)^2} .
    \label{eq:defS2}
\end{equation}
Thus $\e^{-S^{(2)}}=\Tr\rho^2$ is the purity, while $S^{(2)}$ is its negative logarithm. Because the RTE is defined for the full thermal density matrix, it introduces neither a spatial bipartition nor an entangling surface. This convention also fixes the sign of the partition-function difference:
\begin{equation}
    \ln Z(2\beta)-2\ln Z(\beta)=-S^{(2)} .
    \label{eq:RTEsignConvention}
\end{equation}
If $\mathcal L(x,c)$ denotes the universal contribution to $\ln Z$, the corresponding entropy scaling function is $2\mathcal L(x,c)-\mathcal L(x,2c)$. In the residual-free-energy convention $\mathcal F_{\rm crit}=-\mathcal L$, the same quantity is $\mathcal F_{\rm crit}(x,2c)-2\mathcal F_{\rm crit}(x,c)$. Throughout Appendices~\ref{sec:continuous} and \ref{sec:casimireffect}, $\widetilde S_n$ denotes this RTE scaling function, rather than a contribution to $\ln Z$.

With the convention in Eq.~\eqref{eq:defS2}, the left-hand sides of main-text Eqs.~\eqref{eq:differ} and \eqref{Eq:differ} equal $-S^{(2)}$, whereas their right-hand sides follow the sign convention for $S^{(2)}$. Identifying those equations with the RTE therefore requires one overall sign reversal. All expressions below use $S^{(2)}$ and $\partial_gS^{(2)}$ as defined in Eqs.~\eqref{eq:defS2} and \eqref{eq:exactDRTE2}. A global sign reversal changes the orientation of the DRTE curve, but neither its scaling exponent nor the positions of its zeros.

We compare different system sizes along the thermal trajectory
\begin{equation}
    \beta=cL^{z_\beta},
    \qquad c>0 .
    \label{eq:thermalTrajectory}
\end{equation}
At a continuous quantum critical point, we set $z_\beta=z$, where $z$ is the dynamical critical exponent. At a first-order transition, $z_\beta$ is not a universal exponent; it specifies only how the simulation temperature is reduced as $L$ increases. The thermal rounding is then controlled by $\beta L^d\sim L^{d+z_\beta}$, rather than by a correlation-length exponent. The main text uses $c=\beta/L^{z_\beta}$ as the numerical aspect-ratio parameter. For a relativistic critical point with $z_\beta=1$, the metric-normalized temporal aspect ratio is
\begin{equation}
    \rho_\tau=\frac{v\beta}{L}=vc .
    \label{eq:physicalAspectDefinition}
\end{equation}
The simulations with $\beta=L$ therefore have $c=1$ in lattice units, but they are isotropic in spacetime only if the chosen metric normalization gives $v=1$. Appendix~\ref{sec:casimireffect} discusses this distinction in detail.

We define
\begin{equation}
    V(g)=\partial_g H(g)
    \label{eq:defV}
\end{equation}
as the generalized force conjugate to $g$. At fixed $L$ and $\beta$, differentiating the trace and using cyclicity gives
\begin{equation}
    \partial_g\ln Z(s\beta,g,L)
    =-s\beta\avg{V}{s\beta},
    \qquad s>0 .
    \label{eq:dlnZexact}
\end{equation}
This identity does not require $H$ to commute with $\partial_gH$. Differentiating Eq.~\eqref{eq:defRTE} then gives
\begin{equation}
    \partial_g S^{(n)}
    =\frac{n\beta}{n-1}
    \left[
        \avg{V}{n\beta}-\avg{V}{\beta}
    \right] .
    \label{eq:exactDRTEn}
\end{equation}
For the second RTE,
\begin{equation}
    \partial_g S^{(2)}
    =2\beta
    \left[
        \avg{V}{2\beta}-\avg{V}{\beta}
    \right] .
    \label{eq:exactDRTE2}
\end{equation}
In Eqs.~\eqref{eq:exactDRTEn} and \eqref{eq:exactDRTE2}, the derivative is taken at fixed $\beta$ for each finite system. The trajectory $\beta=cL^{z_\beta}$ is imposed only when comparing different sizes.

The RTE is invariant under a common, $g$-dependent shift of all energy levels. If
\begin{equation}
    Z_{\rm ref}(s\beta,g,L)
    =\e^{-s\beta F_{\rm ref}(g,L)},
    \label{eq:singleExpRef}
\end{equation}
then
\begin{equation}
    \ln Z_{\rm ref}(n\beta)-n\ln Z_{\rm ref}(\beta)=0 .
    \label{eq:refCancellation}
\end{equation}
Equivalently, $S^{(n)}$ is unchanged under $H(g)\mapsto H(g)+F_{\rm ref}(g,L)\mathbf 1$. This cancellation removes the extensive ground-state energy and any common term strictly linear in the temporal circumference. On the flat, spatially periodic geometries considered here, ultraviolet-local bulk counterterms have precisely this form. It does not remove nonlocal finite-temperature terms, finite-size excitation towers, or any contribution with nontrivial dependence on the temporal aspect ratio.

This cancellation does not extend to a multilevel sector. If
\begin{equation}
    Z(s\beta,g,L)
    =\e^{-s\beta F_{\rm ref}(g,L)}\,\mathcal Q_s(g,L),
    \qquad
    \mathcal Q_s=\sum_m\e^{-s c\varepsilon_m},
    \label{eq:multiLevelQ}
\end{equation}
then
\begin{equation}
    S^{(n)}
    =\frac{\ln\mathcal Q_n-n\ln\mathcal Q_1}{1-n} .
    \label{eq:RTEafterReferenceCancellation}
\end{equation}
Because $\mathcal Q_n\neq\mathcal Q_1^n$ in general, the finite-size tower gives a physical contribution to the RTE. The distinction between a common single exponential and a common multilevel tower underlies both the critical Casimir construction and the weakly-first-order analysis below. The RTE subtraction removes the leading analytic free-energy background without eliminating the universal finite-size response.

\section{Continuous quantum transitions: a single critical tower}
\label{sec:continuous}

At a continuous transition, the infrared flow approaches a real critical fixed point. For fixed geometry and boundary conditions, the low-energy spectrum forms a single finite-size critical tower~\cite{Wilson1971Renormalization,fisher1972scaling,sachdev2011quantum,Campostrini2014Quantum}. The main text denotes the leading $n=2$, $c=1$ entropy scaling function by $\widetilde S(x)$. Here we keep the \Renyi{} index and temporal aspect ratio explicit and write $\widetilde S_n(x,c)$. We define
\begin{equation}
    x=(g-\gc)L^{1/\nu},
    \qquad
    \beta=cL^z ,
    \label{eq:continuousVariables}
\end{equation}
where $\nu$ and $z$ are the correlation-length and dynamical exponents. The finite-size levels have the scaling form
\begin{equation}
    E_m(g,L)
    =E_{\rm ref}(g,L)
    +L^{-z}\varepsilon_m
    \left(x,u_1L^{-\omega_1},u_2L^{-\omega_2},\ldots\right)
    +o(L^{-z}) .
    \label{eq:criticalSpectrum}
\end{equation}
Here $E_{\rm ref}$ is a common analytic reference energy, $u_i$ are irrelevant scaling fields, and $\omega_i>0$ are correction-to-scaling exponents. The normalization of the dimensionless levels $\varepsilon_m$ includes the temporal metric factor. For $z=1$, Eq.~\eqref{eq:criticalSpectrum} may equivalently be written as
\begin{equation}
    E_m(g,L)=E_{\rm ref}(g,L)
    +\frac{v}{L}\,\mathcal E_m(x,\ldots)+o(L^{-1}),
    \label{eq:criticalSpectrumVelocity}
\end{equation}
so that the Boltzmann factors depend on $\rho_\tau=v\beta/L$. One may either absorb $v$ into $\varepsilon_m=v\mathcal E_m$ and use $c=\beta/L$, as in the main text, or retain $\mathcal E_m$ and use $\rho_\tau$. These conventions are equivalent but should not be combined within a single formula.

At inverse temperature $s\beta=scL^z$,
\begin{equation}
\begin{split}
    Z(s\beta,g,L)
    &=\e^{-s\beta E_{\rm ref}(g,L)}\mathcal Q_s(x,L),\\
    \mathcal Q_s(x,L)
    &=\sum_m\exp\!\left[-sc\,
      \varepsilon_m\!\left(x,u_1L^{-\omega_1},\ldots\right)
      \right] .
\end{split}
    \label{eq:continuousZfactor}
\end{equation}
The tower factor admits the correction-to-scaling expansion
\begin{equation}
    \mathcal Q_s(x,L)
    =Q_s^{(0)}(x,c)
    +\sum_iL^{-\omega_i}Q_{s,i}^{(1)}(x,c)+\cdots .
    \label{eq:QirrelevantExpansion}
\end{equation}
The common exponential in Eq.~\eqref{eq:continuousZfactor} cancels from the RTE, whereas the critical tower remains. Thus
\begin{equation}
    S^{(n)}(g,L;c)
    =\widetilde S_n(x,c)
    +\sum_iL^{-\omega_i}\widetilde S_{n,i}(x,c)+\cdots,
    \label{eq:continuousSnScaling}
\end{equation}
where
\begin{equation}
    \widetilde S_n(x,c)
    =\frac{\ln Q_n^{(0)}(x,c)-n\ln Q_1^{(0)}(x,c)}{1-n} .
    \label{eq:continuousSnFunction}
\end{equation}
The subscript $s$ denotes the same critical tower evaluated at temporal aspect ratio $sc$; it does not represent $s$ independent copies of the $c$ geometry. Independent copies enter only through $Z(\beta)^n$ in the denominator of the RTE ratio. This distinction is central to the spacetime construction in Fig.~\ref{fig:sc}.

The same spectral representation has a direct probability interpretation. Introduce
\begin{equation}
    p_m(x,c)
    =\frac{\e^{-c\varepsilon_m(x)}}{Q_1^{(0)}(x,c)},
    \qquad \sum_m p_m=1 .
    \label{eq:criticalTowerProbabilities}
\end{equation}
Then
\begin{equation}
    \widetilde S_n(x,c)
    =\frac{1}{1-n}\ln\sum_m p_m(x,c)^n .
    \label{eq:criticalTowerRenyi}
\end{equation}
A common shift $\varepsilon_m\mapsto\varepsilon_m+a(x)$ cancels from Eq.~\eqref{eq:criticalTowerRenyi}; only the scaled gaps and their finite-aspect-ratio weights enter.

For $n=2$,
\begin{equation}
    S^{(2)}(g,L;c)
    =2\ln Q_1^{(0)}(x,c)-\ln Q_2^{(0)}(x,c)
    +\Order{L^{-\omega}} .
    \label{eq:continuousS2}
\end{equation}
Differentiating Eq.~\eqref{eq:continuousSnScaling} at fixed $L$ and $\beta$ gives
\begin{equation}
    \partial_gS^{(n)}
    =L^{1/\nu}\partial_x\widetilde S_n(x,c)
    +\sum_iL^{1/\nu-\omega_i}
      \partial_x\widetilde S_{n,i}(x,c)+\cdots .
    \label{eq:continuousDRTE}
\end{equation}
The leading term in Eq.~\eqref{eq:continuousDRTE} arises from the relevant scaling field. Smooth $g$ dependence of nonuniversal metric factors, including $v(g)$ in $\rho_\tau=v(g)\beta/L$, generates additional $O(L^0)$ terms at fixed $\beta$. These terms are subleading to $L^{1/\nu}$ unless symmetry suppresses the leading derivative. A feature at finite scaling coordinate $x_\ast$ therefore satisfies
\begin{equation}
    g_\ast(L)-\gc=x_\ast L^{-1/\nu}+\cdots,
    \qquad
    \left|\partial_gS^{(n)}\right|_\ast\propto L^{1/\nu} .
    \label{eq:continuousPeakScaling}
\end{equation}
At the bulk critical point,
\begin{equation}
    \partial_gS^{(n)}(\gc,L;c)
    =L^{1/\nu}\partial_x\widetilde S_n(0,c)+\cdots .
    \label{eq:continuousAtGc}
\end{equation}
In general, $\partial_x\widetilde S_n(0,c)$ need not vanish. A DRTE zero at a continuous transition therefore requires an additional symmetry or a stationary point of the scaling function; criticality alone does not enforce it.

Because $S^{(n)}$ is dimensionless, curves at fixed aspect ratio approach a common value at $g=\gc$:
\begin{equation}
    S^{(n)}(\gc,L;c)
    =\widetilde S_n(0,c)+L^{-\omega}\widetilde S_{n,1}(0,c)+\cdots .
    \label{eq:continuousCrossingValue}
\end{equation}
Retaining one irrelevant field for sizes $L$ and $bL$, and assuming $\partial_x\widetilde S_n(0,c)\neq0$, gives the crossing drift
\begin{equation}
    g_\times(L)-\gc
    =C(b)L^{-(1/\nu+\omega)}+\cdots .
    \label{eq:crossingDrift}
\end{equation}
If the leading derivative vanishes, the drift is controlled by higher derivatives or subleading correction fields.

Outside the critical window, the limiting RTE is determined by the low-energy spectrum of the phase. A unique gapped ground state gives
\begin{equation}
    Z(\beta)\simeq\e^{-\beta E_0}
    \quad\Longrightarrow\quad
    S^{(n)}\to0,
    \label{eq:uniqueGappedLimit}
\end{equation}
whereas a thermally unresolved $q$-fold manifold gives
\begin{equation}
    Z(\beta)\simeq q\e^{-\beta E_0}
    \quad\Longrightarrow\quad
    S^{(n)}\to\ln q .
    \label{eq:finiteDegeneracyLimit}
\end{equation}
When a continuous symmetry is spontaneously broken, the Anderson tower of states provides an additional contribution~\cite{Anderson1952,Watanabe2020Counting}. For an $O(N)$ order parameter,
\begin{equation}
    E_\ell-E_0
    =\frac{\ell(\ell+N-2)}{2I_L}+\cdots,
    \qquad I_L\propto L^d .
    \label{eq:rotorSpectrum}
\end{equation}
If $\beta\propto L^z$ with $d>z$, then $\beta/I_L\to0$ and
\begin{equation}
    Z_{\rm rot}(\beta)
    \propto\left(\frac{I_L}{\beta}\right)^{(N-1)/2} .
    \label{eq:rotorPartition}
\end{equation}
Consequently,
\begin{equation}
    S_{\rm rot}^{(n)}
    =\frac{N-1}{2}(d-z)\ln L+\Order{1} .
    \label{eq:rotorRenyi}
\end{equation}
This ordered-phase tower is distinct from the critical tower governing Eqs.~\eqref{eq:continuousSnScaling} and \eqref{eq:continuousDRTE}. The continuous-transition results used in the main text are therefore leading finite-size statements at fixed aspect ratio: the RTE approaches a common crossing value up to corrections from irrelevant fields, and the DRTE scales as $L^{1/\nu}$. A single critical tower does not, by itself, enforce either a zero or an opposite-sign two-lobe profile.

\section{Residual finite-size free energy and the spacetime Casimir picture}
\label{sec:casimireffect}

\subsection{Residual free energy on a finite Euclidean geometry}

Here ``Casimir'' refers to the residual, geometry-dependent free energy of a gapless theory on a finite Euclidean manifold after local analytic contributions have been subtracted~\cite{Casimir1948Attraction,FisherdeGennes1978,Krech1994}. For a geometry $\mathcal G_\lambda$ with boundary or periodicity data $\mathcal B_\lambda$, write
\begin{equation}
    -\ln Z[\mathcal G_\lambda,\mathcal B_\lambda]
    =\mathcal F_{\rm loc}[\mathcal G_\lambda,\mathcal B_\lambda]
    +\mathcal F_{\rm res}[\mathcal G_\lambda,\mathcal B_\lambda] .
    \label{eq:casimirResidualDefinition}
\end{equation}
The local term contains ultraviolet-sensitive integrals of local densities, including bulk-volume contributions and, when present, surface, edge, corner, and counterterm contributions. The residual term is the nonlocal finite-size contribution. In the critical scaling limit, it is universal once the geometry, aspect ratios, boundary conditions, and metric normalization are specified.

Hyperscaling fixes the residual contribution to be $O(1)$ in the finite-aspect-ratio regime~\cite{fisher1972scaling,sachdev2011quantum}. The singular free-energy density obeys
\begin{equation}
    f_{\rm sing}(g,T,L^{-1})
    =b^{-(d+z)}
    f_{\rm sing}\!\left((g-\gc)b^{1/\nu},Tb^z,L^{-1}b\right) .
    \label{eq:casimirHyperscaling}
\end{equation}
Setting $b=L$, $x=(g-\gc)L^{1/\nu}$, and $c=\beta/L^z$ gives
\begin{equation}
    f_{\rm sing}(g,T,L^{-1})
    =L^{-(d+z)}\,\mathfrak f(x,c^{-1}) .
    \label{eq:casimirFreeEnergyDensity}
\end{equation}
The corresponding dimensionless Euclidean free energy is
\begin{equation}
    \beta L^d f_{\rm sing}
    =c\,\mathfrak f(x,c^{-1})
    \equiv\mathcal F_{\rm crit}(x,c)=O(1) .
    \label{eq:casimirOone}
\end{equation}
The anomalous dimension $\eta$ does not enter this power counting: it governs correlation functions, not the scaling dimension of the free-energy density.

Gaplessness is essential. In a unique-ground-state gapped phase, the residual dependence on a macroscopic cycle is exponentially suppressed once the cycle exceeds the correlation length. At criticality, no finite infrared scale cuts off this dependence, and the residual free energy retains universal information about the full Euclidean geometry.

\subsection{Parallel plates and the compact-cycle analogue}

The electromagnetic Casimir effect provides the standard reference point~\cite{Casimir1948Attraction,Plunien1986Casimir,Milton2001Casimir}. For two ideal parallel plates separated by $a$, the normal momentum is quantized as $k_z=n\pi/a$. After the bulk and surface self-energies are subtracted, the residual vacuum energy and force per unit area in three spatial dimensions are
\begin{equation}
    \frac{E_{\rm Cas}(a)}{A}
    =-\frac{\pi^2}{720}\frac{\hbar c_0}{a^3},
    \qquad
    \frac{F_{\rm Cas}(a)}{A}
    =-\frac{\partial}{\partial a}\frac{E_{\rm Cas}(a)}{A}
    =-\frac{\pi^2}{240}\frac{\hbar c_0}{a^4} .
    \label{eq:plateCasimir}
\end{equation}
The mechanical force is obtained by differentiating the residual energy with respect to the plate separation. The RTE probes the same infrared mechanism---the dependence of gapless modes on finite geometry---but involves a finite difference of free energies rather than a mechanical derivative.

A more direct spatial analogue of the full-system thermal ratio is a theory with one periodic spatial cycle. Let $\mathcal M_L$ denote the remaining $d-1$ directions, each of characteristic size $L$, and consider
\begin{equation}
    \mathcal G_{\rm sp}(R)=\mathcal M_L\times S_R^1,
    \qquad
    \phi(u+R)=\phi(u),
    \qquad
    r=R/L .
    \label{eq:compactSpatialGeometry}
\end{equation}
The allowed momenta are $k_u=2\pi m/R$. After local terms are removed, write
\begin{equation}
    -\ln Z_{\rm sp}(g,L,R)
    =\mathcal F_{\rm loc}^{\rm sp}(g,L,R)
    +\mathcal F_{\rm Cas}^{\rm sp}(x,r)+o(1) .
    \label{eq:compactSpatialResidual}
\end{equation}
The analogue of the second-RTE construction compares two independent $R$ cycles with one cycle of circumference $2R$:
\begin{equation}
\begin{split}
    \mathcal S_{2,{\rm Cas}}^{\rm sp}(x,r)
    &=\mathcal F_{\rm Cas}^{\rm sp}(x,2r)
      -2\mathcal F_{\rm Cas}^{\rm sp}(x,r)\\
    &=-\ln\frac{\mathcal Z_{\rm Cas}^{\rm sp}(x,2r)}
      {[\mathcal Z_{\rm Cas}^{\rm sp}(x,r)]^2},
\end{split}
    \label{eq:compactSpatialDifference}
\end{equation}
where $\mathcal Z_{\rm Cas}^{\rm sp}=\exp[-\mathcal F_{\rm Cas}^{\rm sp}]$. A $2R$ circle is not equivalent to an $R$ circle traversed twice. Its mode spacing is $\pi/R$, rather than $2\pi/R$, and it therefore supports additional long-wavelength modes.

The periodic scalar field makes this power counting explicit. For a massless real scalar with one compact direction and $d-1$ noncompact transverse directions,
\begin{equation}
    \frac{E_{\rm Cas}^{\rm per}(R)}{V_\perp}
    =-\frac{\Gamma\!\left(\frac{d+1}{2}\right)\zeta(d+1)}
            {\pi^{(d+1)/2}}\frac{1}{R^d} .
    \label{eq:periodicScalarCasimir}
\end{equation}
The coefficient depends on the field content and boundary conditions, while dimensional analysis fixes the dependence on $R$. For $R\sim L$ and $V_\perp\sim L^{d-1}$, the total residual energy scales as $L^{-1}$. At a relativistic critical point, $\beta\sim L$, and hence $\beta E_{\rm Cas}=O(1)$, consistent with Eq.~\eqref{eq:casimirOone}.

\subsection{Temporal compactification and the RTE finite difference}

For the full quantum system, the Euclidean manifold is $\Sigma_L\times S_\beta^1$. In the critical scaling regime, write
\begin{equation}
    -\ln Z(\beta,g,L)
    =\beta E_{\rm loc}(g,L)
    +\mathcal F_{\rm crit}(x,c)+o(1) .
    \label{eq:temporalResidualDecomposition}
\end{equation}
On a smooth periodic spacetime, the local contribution is linear in the temporal circumference and therefore cancels from the full-system RTE. Equation~\eqref{eq:defS2} then gives the singular residual contribution
\begin{equation}
    S_{\rm sing}^{(2)}(x,c)
    =\mathcal F_{\rm crit}(x,2c)
     -2\mathcal F_{\rm crit}(x,c)
    =-\ln\frac{\mathcal Z_{\rm crit}(x,2c)}
       {\mathcal Z_{\rm crit}(x,c)^2}
    \label{eq:temporalCasimirRTE}
\end{equation}
with $\mathcal Z_{\rm crit}=\exp[-\mathcal F_{\rm crit}]$. More generally,
\begin{equation}
    S_{\rm sing}^{(n)}(x,c)
    =\frac{\mathcal F_{\rm crit}(x,nc)
      -n\mathcal F_{\rm crit}(x,c)}{n-1} .
    \label{eq:temporalCasimirGeneralN}
\end{equation}
Geometrically, the numerator of the purity contains one smooth thermal cylinder of circumference $2\beta$, while the denominator contains two disconnected cylinders of circumference $\beta$. Because the sewing acts on the entire spatial system, it creates neither a spatial entangling surface nor a local codimension-two replica defect. We use ``spacetime Casimir effect'' for the residual free-energy difference generated by this global change of temporal compactification.

\subsection{Critical tower and cancellation of the ground-state contribution}

The same result follows directly from the critical spectrum of Appendix~\ref{sec:continuous}. With
\begin{equation}
    E_m(g,L)=E_{\rm ref}(g,L)+L^{-z}\varepsilon_m(x)+o(L^{-z}),
    \label{eq:casimirCriticalSpectrum}
\end{equation}
we have
\begin{equation}
    \mathcal Z_{\rm crit}(x,c)
    =\sum_m\e^{-c\varepsilon_m(x)},
    \qquad
    \mathcal F_{\rm crit}(x,c)=-\ln\mathcal Z_{\rm crit}(x,c) .
    \label{eq:casimirCriticalPartition}
\end{equation}
Defining $p_m(x,c)=\e^{-c\varepsilon_m}/\mathcal Z_{\rm crit}$ gives
\begin{equation}
    S_{\rm sing}^{(2)}(x,c)
    =-\ln\sum_m p_m(x,c)^2 .
    \label{eq:criticalTowerProbabilityCasimir}
\end{equation}
The residual temporal Casimir term is therefore the second \Renyi{} entropy of the normalized weights over the universal finite-size tower.

The ground-state Casimir energy alone is not the RTE contribution. Decompose the scaled levels as
\begin{equation}
    \varepsilon_m(x)=\varepsilon_0(x)+\delta_m(x),
    \qquad \delta_0=0 .
    \label{eq:groundCasimirSplit}
\end{equation}
Then
\begin{equation}
\begin{split}
    \mathcal Z_{\rm crit}(x,c)
    &=\e^{-c\varepsilon_0(x)}
      \left[1+\sum_{m>0}\e^{-c\delta_m(x)}\right],\\
    \mathcal F_{\rm crit}(x,c)
    &=c\varepsilon_0(x)
      -\ln\left[1+\sum_{m>0}\e^{-c\delta_m(x)}\right] .
\end{split}
    \label{eq:groundCasimirFreeEnergy}
\end{equation}
The term $c\varepsilon_0$ is linear in the temporal circumference and cancels identically in the RTE finite difference:
\begin{equation}
    2c\varepsilon_0-2(c\varepsilon_0)=0 .
    \label{eq:groundCasimirCancel}
\end{equation}
For a nondegenerate ground state, the surviving contribution is
\begin{equation}
\begin{split}
    S_{\rm sing}^{(2)}(x,c)
    =&\;2\ln\left[1+\sum_{m>0}\e^{-c\delta_m(x)}\right]\\
     &-\ln\left[1+\sum_{m>0}\e^{-2c\delta_m(x)}\right] .
\end{split}
    \label{eq:excitedTowerCasimir}
\end{equation}
The scaled ground-state Casimir term exhibits the same $O(1)$ power counting as the residual free energy, but cancels from the RTE. The surviving finite difference is governed by the finite-aspect-ratio occupation of the excitation tower. A thermally unresolved ground-state degeneracy is incorporated by replacing the leading $1$ in Eq.~\eqref{eq:excitedTowerCasimir} with the corresponding degeneracy.

\subsection{Matsubara scale, temporal aspect ratio, and the spacetime Casimir picture}
\label{sec:matsubaraDetailed}

We now specialize to the $(2+1)$-dimensional geometries used in the simulations. The full spatial manifold is the periodic $L\times L$ torus,
\begin{equation}
    \Sigma_L=T_L^2,
\end{equation}
and the thermal partition function has the Hamiltonian representation
\begin{equation}
    Z_{\Sigma}(\beta,g,L)
    =\Tr_{\mathcal H(\Sigma_L)}
      \e^{-\beta H_{\Sigma}(g,L)} .
    \label{eq:originalHamiltonianChannel}
\end{equation}
Equivalently, it is the Euclidean path integral on
\begin{equation}
    T_L^2\times S_\beta^1 .
    \label{eq:euclideanThreeTorus}
\end{equation}
The Hamiltonian trace and the Euclidean path integral are two representations of the same partition function.

At fixed $L$ and $g$, varying $\beta$ in Eq.~\eqref{eq:originalHamiltonianChannel} does not change the eigenvalues of $H_{\Sigma}(g,L)$; it changes only their Boltzmann weights. At a relativistic critical point, the low-energy levels take the form
\begin{equation}
    E_m(g,L)
    =E_{\rm ref}(g,L)
     +\frac{v}{L}\,\mathcal E_m(x)
     +o(L^{-1}),
    \qquad
    x=(g-\gc)L^{1/\nu},
    \label{eq:matsubaraCriticalSpectrumVelocity}
\end{equation}
where $v$ is the emergent velocity and $\mathcal E_m$ denotes dimensionless torus-spectrum data for the chosen geometry and boundary conditions. This is Eq.~\eqref{eq:casimirCriticalSpectrum} with $z=1$ and $\varepsilon_m=v\mathcal E_m$.

To compare imaginary time with a spatial length, the velocity must be included. Introducing the length coordinate
\begin{equation}
    y=v\tau,
\end{equation}
the circumference of the Euclidean-time cycle is
\begin{equation}
    L_\tau=v\beta,
    \qquad
    \rho_\tau=\frac{L_\tau}{L}=\frac{v\beta}{L} .
    \label{eq:physicalTemporalAspectRatio}
\end{equation}
The infrared critical theory therefore lives on a three-torus with cycle lengths $L$, $L$, and $L_\tau$. The simulation convention
\begin{equation}
    \beta=cL
    \label{eq:qmcAspectConvention}
\end{equation}
keeps $\rho_\tau=vc$ fixed at criticality. The simulations reported in the main text use $\beta=L$, so $c=1$ in lattice units and $\rho_\tau=v$. This is a valid fixed-aspect-ratio trajectory, but it is not necessarily an isotropic three-torus. Isotropy in the continuum metric requires $L_\tau=L$, or $\beta=L/v$. Thus $c$ is the numerical time-to-space ratio used in the main text, whereas universal torus amplitudes depend on the metric-normalized ratio $\rho_\tau$.

The path integral provides the complementary Matsubara description. A bosonic critical field is periodic around the thermal circle and admits the expansion
\begin{equation}
    \phi(\tau,\bm r)
    =\sum_{n\in\mathbb Z}
      \e^{-\mathrm i\omega_n\tau}\phi_n(\bm r),
    \qquad
    \omega_n=\frac{2\pi n}{\beta} .
    \label{eq:matsubaraModeExpansion}
\end{equation}
In the length coordinate $y=v\tau$, the corresponding temporal wave number is
\begin{equation}
    k_{\tau,n}=\frac{\omega_n}{v}
    =\frac{2\pi n}{v\beta}
    =\frac{2\pi n}{L_\tau} .
    \label{eq:matsubaraTemporalMomentum}
\end{equation}
Finite $\beta$ therefore discretizes the frequency direction. The static $n=0$ sector is present for any circumference, while the smallest nonzero bosonic frequency is $2\pi/\beta$. Shortening the time circle raises this infrared scale and precludes nonconstant temporal variations whose wavelength would exceed the circumference. This kinematic statement remains valid in an interacting critical theory, although the partition function need not factorize into independent oscillator modes.

For the second RTE, the comparison is particularly transparent. On a circle of circumference $\beta$,
\begin{equation}
    \omega_m^{(\beta)}=\frac{2\pi m}{\beta},
\end{equation}
whereas on a circle of circumference $2\beta$,
\begin{equation}
    \omega_n^{(2\beta)}=\frac{\pi n}{\beta} .
    \label{eq:betaTwoBetaMatsubara}
\end{equation}
Even values $n=2m$ reproduce the $\beta$-circle frequencies, whereas odd values $n=2m+1$ fall between them. The longer circle therefore admits additional nonzero frequencies between those of the $\beta$ circle. Both spectra remain unbounded in the ultraviolet: doubling the circumference does not enlarge a microscopic bandwidth, but instead refines the infrared frequency grid. Moreover, $Z(\beta)^2$ describes two disconnected copies of the $\beta$ geometry, not a single circle with twice the mode multiplicity. One $2\beta$ circle and two independent $\beta$ circles therefore have different residual free energies.

Figure~\ref{fig:sc} illustrates this geometric comparison. Panel (a) represents the single connected Euclidean cylinder contributing $Z(2\beta)$, and panel (b) the two disconnected cylinders contributing $Z(\beta)^2$. The label $A$ denotes the complete spatial system; one spatial direction of the actual $T_L^2$ geometry is suppressed. The arrows indicate periodic tracing in imaginary time.

The red curves are schematic long-wavelength critical fluctuations, not individual many-body eigenstates, and neither their number nor their thickness is quantitative. The thick, slowly varying curve drawn only on the $2\beta$ cylinder represents a fluctuation in the additional soft nonzero temporal sector allowed by the smaller spacing in Eq.~\eqref{eq:betaTwoBetaMatsubara}. Accordingly, the term ``sieve'' in the main text refers to the infrared quantization of nonzero temporal wavelengths. The ``energy cost'' in the caption denotes the change in the dimensionless residual free energy $-\ln Z$, not an eigenvalue assigned to a particular curve. Varying $\beta$ reweights the fixed Hamiltonian spectrum; it does not remove selected microscopic modes.

A dual transfer-matrix channel makes the spatial Casimir analogy explicit. After the rescaling $y=v\tau$, the infrared path integral is defined on the three-torus $(L,L,L_\tau)$. One may slice this manifold along a spatial direction rather than along $y$. With the dual Hamiltonian normalized in physical energy units,
\begin{equation}
    Z_{\Sigma}(\beta,g_c,L)
    =\Tr_{\mathcal H(T^2_{L_\tau,L})}
      \exp\!\left[-\frac{L}{v}
      H_{\rm dual}(L_\tau,L)\right] .
    \label{eq:dualTransferChannel}
\end{equation}
Equation~\eqref{eq:dualTransferChannel} is simply another transfer-matrix decomposition of the same critical path integral; it does not introduce a new microscopic model.

For a $(2+1)$-dimensional relativistic critical theory on a spatial torus with circumferences $L_x$ and $L_y$, the finite-size levels have the general form
\begin{equation}
    E_m(L_x,L_y)-E_{\rm ref}(L_x,L_y)
    =\frac{v}{L_x}\,
      \mathcal E_m\!\left(\frac{L_y}{L_x}\right)
      +o(L_x^{-1}) .
    \label{eq:twoDimensionalTorusSpectrum}
\end{equation}
Applying this to the dual channel with $L_x=L_\tau=v\beta$ and $L_y=L$ gives
\begin{equation}
\begin{split}
    E_m^{\rm dual}-E_0^{\rm dual}
    =&\;\frac{1}{\beta}
    \left[
      \mathcal E_m^{\rm dual}(\rho_\tau^{-1})
      -\mathcal E_0^{\rm dual}(\rho_\tau^{-1})
    \right]\\
    &+o(\beta^{-1}) .
\end{split}
    \label{eq:dualChannelGapScale}
\end{equation}
The dual finite-size gaps and the Matsubara spacing are therefore set by the same inverse-time scale, $1/\beta$, up to dimensionless geometry-dependent factors. The kinematic factor $2\pi$ in the Matsubara frequencies should not be identified with the interacting torus spectrum. In particular, $E_m-E_0=(2\pi v/L)\Delta_m$ is specific to a $(1+1)$-dimensional CFT quantized on a circle. In $(2+1)$ dimensions, the state-operator correspondence maps local operators to states on $S^2$, not to the spectrum on $T^2$. Torus levels are universal, but depend on the torus shape and boundary conditions.

The Hamiltonian, Matsubara, and dual-channel pictures are mutually consistent. In the Hamiltonian channel, $\beta$ sets the occupation of the fixed finite-size levels in Eq.~\eqref{eq:matsubaraCriticalSpectrumVelocity}. In the Matsubara representation, it fixes the temporal frequency spacing. In the dual channel, $L_\tau=v\beta$ becomes a compact spatial circumference and produces gaps of order $1/\beta$. These are three representations of the same geometric dependence of the critical partition function, not separate additive effects.

With the entropy convention of Eq.~\eqref{eq:defS2}, the geometry used in the main text gives
\begin{equation}
    S^{(2)}(\beta,g,L)
    =-\ln\frac{Z_{\Sigma}(2\beta,g,L)}
                    {Z_{\Sigma}(\beta,g,L)^2} .
    \label{eq:matsubaraRTEratio}
\end{equation}
After cancellation of the local terms linear in the temporal circumference, the critical residual contribution is
\begin{equation}
    S_{\rm sing}^{(2)}(x,c)
    =\mathcal F_{\rm crit}(x,2c)
     -2\mathcal F_{\rm crit}(x,c) .
    \label{eq:matsubaraCasimirDifference}
\end{equation}
For the data in the main text, $c=1$ and hence
\begin{equation}
    S_{\rm sing}^{(2)}(x,1)
    =\mathcal F_{\rm crit}(x,2)
     -2\mathcal F_{\rm crit}(x,1) .
    \label{eq:mainTextCasimirAspect}
\end{equation}
In metric-normalized variables, the two terms evaluate the same critical theory at temporal aspect ratios $2v$ and $v$, respectively. The spacetime Casimir contribution in this work is the finite difference in Eq.~\eqref{eq:matsubaraCasimirDifference}, not a mechanical derivative with respect to a plate separation.

The order of limits is essential. At criticality,
\begin{equation}
    \Delta_L=E_1-E_0
    =\frac{v}{L}
      [\mathcal E_1(0)-\mathcal E_0(0)]
      +o(L^{-1}),
    \label{eq:criticalGapAspect}
\end{equation}
so along $\beta=cL$,
\begin{equation}
    \beta\Delta_L
    =cv[\mathcal E_1(0)-\mathcal E_0(0)]+o(1) .
    \label{eq:finiteAspectBetaGap}
\end{equation}
The finite-aspect-ratio limit
\begin{equation}
    L\rightarrow\infty,
    \qquad
    \beta=cL,
    \qquad
    c=O(1),
    \label{eq:finiteAspectOrderOfLimits}
\end{equation}
therefore retains a finite occupation of the universal critical tower. By contrast, taking $\beta\rightarrow\infty$ at fixed $L$ projects onto a unique ground state and gives $S^{(2)}\rightarrow0$. In a unique-ground-state gapped phase with thermodynamic gap $\Delta>0$, the trajectory $\beta=cL$ yields $\beta\Delta\rightarrow\infty$, so the critical residual contribution is exponentially suppressed. This statement applies to the critical Casimir sector: a finite ground-state degeneracy contributes $\ln q$, and a phase with spontaneously broken continuous symmetry may retain the Anderson-tower term discussed in Appendix~\ref{sec:continuous}.

In this sense, the spacetime Casimir picture is a controlled infrared interpretation of the RTE finite difference: it isolates the universal response of a gapless finite-size spectrum to temporal compactification, without assigning independent microscopic modes to the schematic curves in Fig.~\ref{fig:sc}.

\section{First-order quantum transitions: two gapped branches}
\label{sec:firstorder}

\subsection{Two-branch partition function}

At a zero-temperature first-order transition, two thermodynamic branches cross in energy density while retaining different generalized-force densities. The standard formula in the main text applies when four conditions hold: the coexistence window is dominated by two branches; each branch has a finite, $s$-independent multiplicity; intrabranch excitations satisfy $\beta\Delta_\alpha\gg1$; and tunneling and interface sectors are exponentially suppressed. We write
\begin{equation}
    E_\alpha(g,L)
    =L^d e_\alpha(g)+E_\alpha^{\rm sub}(g,L),
    \qquad \alpha=1,2,
    \label{eq:FObranchEnergies}
\end{equation}
with
\begin{equation}
    e_1(\gc)=e_2(\gc),
    \qquad
    \ell=\partial_g[e_2(g)-e_1(g)]_{g=\gc}\neq0 .
    \label{eq:appLatentForce}
\end{equation}
The coefficient $\ell$ is the discontinuity in the generalized-force density. Because each branch remains gapped, its low-temperature contribution in the coexistence window is
\begin{equation}
    Z_\alpha(s\beta)
    =d_\alpha\e^{-s\beta E_\alpha(g,L)}
    \left[1+\Order{\e^{-s\beta\Delta_\alpha}}\right],
    \label{eq:appRestrictedBranchZ}
\end{equation}
where $d_\alpha$ is a finite multiplicity and $\Delta_\alpha>0$ is the intrabranch gap. Up to exponentially small interface, tunneling, and intrabranch corrections,
\begin{equation}
    Z(s\beta,g,L)
    \simeq d_1\e^{-s\beta E_1(g,L)}
    +d_2\e^{-s\beta E_2(g,L)} .
    \label{eq:appFOtwoBranchZ}
\end{equation}
We define
\begin{equation}
    \Delta E(g,L)=E_2(g,L)-E_1(g,L),
    \qquad
    y=\beta\Delta E(g,L) .
    \label{eq:appFOy}
\end{equation}
After factoring out branch 1,
\begin{equation}
    Z(s\beta)=\e^{-s\beta E_1}
    \left[d_1+d_2\e^{-sy}\right] .
    \label{eq:appFOfactoredZ}
\end{equation}
The common exponential cancels from the RTE, giving
\begin{equation}
    S_{\FO}^{(n)}(y)
    =\frac{
      \ln\left(d_1+d_2\e^{-ny}\right)
      -n\ln\left(d_1+d_2\e^{-y}\right)
    }{1-n} .
    \label{eq:FOSnGeneral}
\end{equation}
At the branch-energy crossing,
\begin{equation}
    S_{\FO}^{(n)}(0)=\ln(d_1+d_2) .
    \label{eq:appFOSatCrossing}
\end{equation}
The unshifted RTE of Eq.~\eqref{eq:defRTE} is therefore finite at coexistence and equals $\ln2$ for two nondegenerate branches. The sign change occurs in the derivative, not in the entropy itself. A coupling-independent plotting offset leaves the DRTE unchanged, but the shifted quantity is no longer the absolute RTE defined in Eq.~\eqref{eq:defRTE}.

\subsection{Occupation identity, zero crossing, and finite-size scaling}

The occupation probability of branch 2 in the $s\beta$ ensemble is
\begin{equation}
    p_s(y)
    =\frac{d_2\e^{-sy}}{d_1+d_2\e^{-sy}},
    \qquad s>0 .
    \label{eq:appFOoccupation}
\end{equation}
Differentiating Eq.~\eqref{eq:FOSnGeneral} gives
\begin{equation}
    \partial_gS_{\FO}^{(n)}
    =\frac{n}{n-1}\,
    \partial_gy\,[p_n(y)-p_1(y)] .
    \label{eq:appFODRTEgeneral}
\end{equation}
For $n=2$,
\begin{equation}
    \partial_gS_{\FO}^{(2)}
    =2\beta\partial_g\Delta E(g,L)
    [p_2(y)-p_1(y)] .
    \label{eq:appFODRTE2exact}
\end{equation}
Using Eq.~\eqref{eq:FObranchEnergies},
\begin{equation}
    \partial_gS_{\FO}^{(2)}
    =2\beta L^d\Delta e'(g)
    [p_2(y)-p_1(y)]+\cdots,
    \qquad \Delta e=e_2-e_1 .
    \label{eq:FODRTE2leading}
\end{equation}
At $\Delta E=0$,
\begin{equation}
    p_n(0)=p_1(0)=\frac{d_2}{d_1+d_2},
    \qquad
    \partial_gS_{\FO}^{(n)}\big|_{\Delta E=0}=0 .
    \label{eq:appFOzeroCrossing}
\end{equation}
At the branch-energy crossing, the occupations in the $n\beta$ and $\beta$ ensembles coincide even when $d_1\neq d_2$, provided the multiplicities are finite constants and the intrabranch spectra are frozen. This should be distinguished from equal occupation within a single $s\beta$ ensemble, which occurs at $sy=\ln(d_2/d_1)$ and therefore depends on $s$. For $y>0$, branch 2 lies higher in energy and is more strongly suppressed in the colder $n\beta$ ensemble, so $p_n<p_1$; for $y<0$, the inequality reverses. Because $p_n-p_1$ also vanishes in both pure-phase limits, the DRTE has two lobes of opposite sign. Their orientation depends on the branch labels and on the sign of $\partial_g\Delta E$. Equal multiplicities give an antisymmetric leading line shape, whereas unequal multiplicities skew the lobes without removing the zero at $y=0$.

Let $g_0(L)$ denote the finite-size branch crossing,
\begin{equation}
    \Delta E[g_0(L),L]=0,
    \qquad g_0(L)\to\gc .
    \label{eq:FOg0}
\end{equation}
Near this point,
\begin{equation}
    \Delta E(g,L)
    =L^d\ell_L[g-g_0(L)]+\cdots,
    \qquad \ell_L\to\ell .
    \label{eq:FOdeltaEExpansion}
\end{equation}
Along $\beta=cL^{z_\beta}$,
\begin{equation}
    y=c\ell_L[g-g_0(L)]L^{d+z_\beta}+\cdots .
    \label{eq:FOscalingVariable}
\end{equation}
The thermal coexistence window is defined by $|y|=O(1)$, which gives
\begin{equation}
    \delta g_{\FO}\sim L^{-(d+z_\beta)},
    \qquad
    \left|\partial_gS_{\FO}^{(n)}\right|_{\rm peak}
    \sim L^{d+z_\beta} .
    \label{eq:appFOwidthHeight}
\end{equation}
The leading finite-size form is
\begin{equation}
    \partial_gS_{\FO}^{(n)}
    =L^{d+z_\beta}
    \mathcal F_n\!\left([g-g_0(L)]L^{d+z_\beta};d_2/d_1\right)
    +\cdots .
    \label{eq:FOFSS}
\end{equation}
The notation $1/\nu_{\FO}=d+z_\beta$ is merely a convenient parametrization of discontinuity rounding; it does not imply a diverging correlation length or a universal first-order dynamical exponent.

Subextensive branch energies determine the drift of the finite-size crossing:
\begin{equation}
    g_0(L)-\gc
    =-\frac{E_2^{\rm sub}(\gc,L)-E_1^{\rm sub}(\gc,L)}{L^d\ell}
    +\cdots .
    \label{eq:FOcrossingShift}
\end{equation}
This drift is distinct from the narrower thermal rounding width in Eq.~\eqref{eq:appFOwidthHeight}. The exact finite-size zero therefore occurs at $g_0(L)$ and need not coincide with the thermodynamic coupling $\gc$ at finite $L$. Constant multiplicities modify the RTE profile but do not shift the zero away from $\Delta E=0$. If either branch retains an aspect-ratio-dependent low-energy tower, its contribution cannot be reduced to a constant; Appendix~\ref{sec:weaklyfirstorder} treats that weakly-first-order regime.

\subsection{Finite-size tunneling}

If the two branches share the same exact quantum numbers, finite-size tunneling can produce an avoided crossing. In the two-branch subspace,
\begin{equation}
    H_{\rm eff}
    =\bar E\,\mathbf 1+\frac{\Delta E}{2}\sigma^z+t_L\sigma^x,
    \qquad
    \Omega_L=\sqrt{(\Delta E/2)^2+t_L^2} .
    \label{eq:FOtwoLevelHamiltonian}
\end{equation}
The corresponding partition function is
\begin{equation}
    Z(s\beta)=2\e^{-s\beta\bar E}\cosh(s\beta\Omega_L) .
    \label{eq:FOtwoLevelZ}
\end{equation}
At a first-order transition, $t_L$ is typically exponentially small in a positive power of $L$. Since $\beta$ grows only algebraically, $\beta t_L\to0$, and the incoherent two-branch result is recovered, with corrections of order $(\beta t_L)^2$ near coexistence. A smooth $g$ dependence of $t_L$ produces corrections of the same exponential order. If the branches belong to different exact symmetry sectors, $t_L=0$ identically. Interface states and tunneling instantons are likewise exponentially suppressed in the asymptotic two-branch regime. Under these conditions, the zero crossing, the $L^{d+z_\beta}$ peak scale, and the $L^{-(d+z_\beta)}$ width all follow from the same occupation identity.

\section{weakly first-order transitions: coexistence and pseudocritical towers}
\label{sec:weaklyfirstorder}

A weakly first-order transition is characterized by a finite but potentially very large crossover length $\xi_\star$. For $L\lesssim\xi_\star$, the RG flow may linger near a real multicritical point or a pair of complex fixed points, producing walking or pseudocritical scaling with effective exponents $\nu_\star$ and $z_\star$, even though the thermodynamic transition is first order~\cite{gorbenko2018walking,gorbenko2018walking2}. We define
\begin{equation}
    x=(g-g_\star)L^{1/\nu_\star},
    \qquad
    \beta=cL^{z_\beta},
    \qquad z_\beta=z_\star,
    \label{eq:WFOvariables}
\end{equation}
where $g_\star$ denotes the reference coupling about which the walking regime is organized. Within this finite-size window, the symbols $g_c$, $\nu$, and $z$ used in the main-text pseudocritical ansatz correspond to $g_\star$, $\nu_\star$, and $z_\star$, respectively. We retain $z_\beta$ to emphasize that the thermal trajectory is an input; choosing $z_\beta=z_\star$ keeps the pseudocritical tower at fixed temporal aspect ratio. The finite-size trace then contains two macroscopic branches, each dressed by a nontrivial low-energy tower.

\subsection{Branch-resolved finite-size trace}

Let $\alpha=1,2$ label the branches and $m$ the levels within each branch. In the walking regime, we write
\begin{equation}
    E_{\alpha m}(g,L)
    =L^d e_\alpha(g)+L^{-z_\star}\varepsilon_{\alpha m}(x)+\cdots .
    \label{eq:appWFOspectrum}
\end{equation}
At inverse temperature $s\beta=scL^{z_\beta}=scL^{z_\star}$,
\begin{equation}
\begin{split}
    Z_\alpha(s\beta)
    &=\e^{-s\beta L^d e_\alpha(g)}Q_{\alpha,s}(x,c,L),\\
    Q_{\alpha,s}(x,c,L)
    &\equiv Q_\alpha(sc,x)
     =\sum_m\e^{-sc\varepsilon_{\alpha m}(x)}+\cdots .
\end{split}
    \label{eq:appWFOrestrictedZ}
\end{equation}
The full low-energy trace is
\begin{equation}
    Z(s\beta)
    \simeq
    \e^{-s\beta L^d e_1(g)}Q_{1,s}
    +\e^{-s\beta L^d e_2(g)}Q_{2,s} .
    \label{eq:appWFOtwoTowerZ}
\end{equation}
Unlike in the asymptotic first-order regime, $Q_{\alpha,s}$ are generally $O(1)$ functions of both $x$ and the temporal aspect ratio $sc$. They cannot be replaced by constant multiplicities.

\subsection{Branch-symmetric common and relative sectors}

A branch-symmetric decomposition follows from
\begin{equation}
\begin{gathered}
    \bar e=\frac{e_1+e_2}{2},
    \qquad
    \Delta e=e_2-e_1,\\
    Q_{{\rm com},s}=(Q_{1,s}Q_{2,s})^{1/2},
    \qquad
    \delta_s=\frac12\ln\frac{Q_{1,s}}{Q_{2,s}} .
\end{gathered}
    \label{eq:WFOsymmetricDefs}
\end{equation}
Then Eq.~\eqref{eq:appWFOtwoTowerZ} becomes
\begin{equation}
    Z(s\beta)
    =2\e^{-s\beta L^d\bar e}\,Q_{{\rm com},s}\cosh Y_s+\cdots,
    \qquad
    Y_s=\frac{sc}{2}L^{d+z_\beta}\Delta e+\delta_s .
    \label{eq:WFOsymmetricZ}
\end{equation}
Interchanging the two branches sends $Y_s\mapsto-Y_s$ and leaves $Q_{{\rm com},s}$ invariant. The decomposition therefore does not privilege either branch.

To leading order, the second RTE is
\begin{equation}
\begin{split}
    S^{(2)}
    =&\;A_{\rm com}(x;c,L)
    +2\ln[2\cosh Y_1]-\ln[2\cosh Y_2]+\cdots,\\
    A_{\rm com}(x;c,L)
    =&\;2\ln Q_{{\rm com},1}-\ln Q_{{\rm com},2} .
\end{split}
    \label{eq:WFOcommonRTE}
\end{equation}
The mean extensive branch energy $\bar e$ cancels exactly. By contrast, $A_{\rm com}$ is the RTE of the common multilevel tower and is generally a physical contribution. It may be neglected only when it is $g$ independent over the fitted window, when symmetry forces its derivative to vanish, or when a background-subtracted observable is explicitly defined. A normalization convention alone cannot remove a $g$-dependent critical tower from the measured RTE.

The occupation of branch 2 in the $s\beta$ ensemble is
\begin{equation}
    P_{2,s}=\frac{1}{1+\e^{2Y_s}},
    \qquad
    P_{1,s}=1-P_{2,s} .
    \label{eq:WFOoccupations}
\end{equation}
Differentiating Eq.~\eqref{eq:WFOcommonRTE} yields
\begin{equation}
    \partial_gS^{(2)}
    =\partial_gA_{\rm com}
    +2\tanh(Y_1)\partial_gY_1
    -\tanh(Y_2)\partial_gY_2+\cdots,
    \label{eq:WFOexactSymmetricDRTE}
\end{equation}
with
\begin{equation}
    \partial_gY_s
    =\frac{sc}{2}L^{d+z_\beta}\Delta e'(g)
    +\frac{L^{1/\nu_\star}}{2}
      \partial_x\ln\frac{Q_{1,s}}{Q_{2,s}}+\cdots .
    \label{eq:WFOdYsSymmetric}
\end{equation}
Direct differentiation of Eq.~\eqref{eq:appWFOtwoTowerZ} gives the equivalent form
\begin{equation}
\begin{split}
    \partial_gS^{(2)}
    =&\;2cL^{d+z_\beta}\Delta e'(g)
       [P_{2,2}-P_{2,1}]\\
     &+L^{1/\nu_\star}
       \left[
       2\sum_{\alpha=1}^2P_{\alpha,1}\partial_x\ln Q_{\alpha,1}
       -\sum_{\alpha=1}^2P_{\alpha,2}\partial_x\ln Q_{\alpha,2}
       \right]+\cdots .
\end{split}
    \label{eq:WFOgeneralDRTE}
\end{equation}
Equation~\eqref{eq:WFOgeneralDRTE} is the general leading-order result in the weakly-first-order regime. The first line describes macroscopic transfer of weight between the two branches. The second contains both the response of the common critical tower and relative tower corrections. It scales as $L^{1/\nu_\star}$ within the walking regime and need not have the same line shape as the occupation-transfer term.

\subsection{Single-mode reduction and the main-text form}

When a single effective splitting controls the relative branch weight, Eq.~\eqref{eq:WFOgeneralDRTE} reduces to a compact form. We parameterize
\begin{equation}
    2Y_s=sX+\lambda+\cdots,
    \qquad
    X=cL^{d+z_\beta}\Delta e(g)+c\phi(x),
    \label{eq:WFOsingleMode}
\end{equation}
where $\lambda$ is a smooth, approximately $s$-independent relative multiplicity offset and $\phi(x)$ is the leading pseudocritical correction to the branch splitting. Then
\begin{equation}
    P_{2,s}(X)=\frac{1}{1+\e^{sX+\lambda}} .
    \label{eq:WFOsingleOccupation}
\end{equation}
Substituting Eq.~\eqref{eq:WFOsingleMode} into Eq.~\eqref{eq:WFOexactSymmetricDRTE} gives
\begin{equation}
\begin{split}
    \partial_gS^{(2)}
    =&\;L^{1/\nu_\star}\partial_xA_{\rm com}(x;c,L)\\
     &+\left[
       2cL^{d+z_\beta}\Delta e'(g)
       +2cL^{1/\nu_\star}\phi'(x)
       \right]
       [P_{2,2}(X)-P_{2,1}(X)]
       +\cdots .
\end{split}
    \label{eq:WFOmainTextForm}
\end{equation}
The compact expression in the main text follows when the first line is negligible, forbidden by symmetry, or removed through an explicitly defined smooth-background subtraction. The correspondence between Eq.~\eqref{eq:WFOmainTextForm} and main-text Eq.~\eqref{eq:WFOmainTextForm1} is
\begin{equation}
\begin{gathered}
    y_{\rm main}\longleftrightarrow X,
    \qquad
    p_s(y_{\rm main})\longleftrightarrow P_{2,s}(X),\\
    \widetilde S'_{\rm main}(x)
    \longleftrightarrow 2c\,\phi'(x),
    \qquad
    z_\beta=z_\star .
\end{gathered}
    \label{eq:WFOmainTextDictionary}
\end{equation}
Thus the variable $y$ in the main-text occupation functions corresponds to the effective relative log-weight $X$, which includes both the macroscopic branch splitting and its leading pseudocritical correction. The two powers of $L$ in the second line of Eq.~\eqref{eq:WFOmainTextForm} arise from differentiating this same relative splitting. The first line is instead the independent RTE response of the common critical tower, a physical $g$-dependent multilevel contribution that cannot be removed by normalization alone.

For equal effective multiplicities, $\lambda=0$, the occupation difference is
\begin{equation}
    P_{2,2}(X)-P_{2,1}(X)
    =\frac{1}{1+\e^{2X}}-\frac{1}{1+\e^X}
    =\frac{\e^X(1-\e^X)}{(1+\e^X)(1+\e^{2X})} .
    \label{eq:WFOoccupationDifference}
\end{equation}
This function vanishes as $X\to\pm\infty$, changes sign at $X=0$, and has one extremum on each side. Unequal but $s$-independent multiplicities preserve the zero at $X=0$ while making the two lobes asymmetric.

\subsection{Zero crossing, crossover scales, and asymptotic recovery}

We define the finite-size relative-coexistence point by
\begin{equation}
    X[g_0(L),L]=0 .
    \label{eq:WFOcoexistenceCondition}
\end{equation}
At this point $P_{2,2}=P_{2,1}$, and the occupation-transfer term in Eq.~\eqref{eq:WFOmainTextForm} vanishes. The total DRTE is
\begin{equation}
    \partial_gS^{(2)}\big|_{g_0(L)}
    =L^{1/\nu_\star}\partial_xA_{\rm com}(x_0;c,L)+\cdots .
    \label{eq:WFOzeroCondition}
\end{equation}
The occupation-transfer component therefore vanishes exactly at relative coexistence. At finite $L$, the full DRTE has an exact zero at the same point only if $\partial_xA_{\rm com}(x_0;c,L)=0$.

If this derivative is parametrically smaller than the occupation-transfer slope, it produces only a subleading displacement of the zero. If the relative amplitudes have different $s$ dependence, the condition $P_{2,2}=P_{2,1}$ may itself shift away from $X=0$. In the asymptotic gapped two-branch regime, $Q_{\alpha,s}$ approach constants, $\partial_xA_{\rm com}\to0$, and the exact first-order zero of Appendix~\ref{sec:firstorder} is recovered.

Within the single-mode regime, the two contributions to the relative prefactor define the amplitude scales
\begin{equation}
    A_{\FO}(L)=2c|\Delta e'(g_0)|L^{d+z_\beta},
    \qquad
    A_\star(L)=2c|\phi'(x_0)|L^{1/\nu_\star} .
    \label{eq:WFOamplitudes}
\end{equation}
If $d+z_\beta>1/\nu_\star$, their crossover estimate is
\begin{equation}
    L_\times
    \sim
    \left[
      \frac{|\phi'(x_0)|}{|\Delta e'(g_0)|}
    \right]^{1/(d+z_\beta-1/\nu_\star)} .
    \label{eq:WFOcrossoverLength}
\end{equation}
For $L\ll L_\times$, the lobe amplitude can grow as $L^{1/\nu_\star}$ even though coexistence produces its sign-changing shape. For $L\gg L_\times$, the branch-slope term proportional to $L^{d+z_\beta}$ dominates. A common-tower contribution of order $L^{1/\nu_\star}$ can add a smooth offset or a single-peak component and must be included when interpreting the full line shape.

Expanding Eq.~\eqref{eq:WFOsingleMode} near $g_0(L)$ gives
\begin{equation}
    X\simeq c\left[
       \Delta e'(g_0)L^{d+z_\beta}
       +\phi'(x_0)L^{1/\nu_\star}
    \right](g-g_0) .
    \label{eq:WFOwidthExpansion}
\end{equation}
The condition $|X|=O(1)$ then gives
\begin{equation}
    \delta g_{\WFO}\sim
    \begin{cases}
      L^{-1/\nu_\star}, & L\ll L_\times,\\[2pt]
      L^{-(d+z_\beta)}, & L\gg L_\times,
    \end{cases}
    \label{eq:WFOwidthCrossover}
\end{equation}
provided the walking description remains valid. For $L\gg\xi_\star$, the pseudocritical tower eventually freezes and the first-order scaling of Appendix~\ref{sec:firstorder} is recovered. An $L^{1/\nu_\star}$ collapse over accessible sizes therefore identifies a pseudocritical regime, but does not by itself determine the thermodynamic order. The occupation-transfer contribution and its crossover must be examined concurrently.

\section{Unified interpretation, diagnostic criteria, and range of validity}
\label{sec:summary}

Equation~\eqref{eq:exactDRTE2} compares the generalized force of the same finite-size Hamiltonian at inverse temperatures $\beta$ and $2\beta$. Throughout these appendices, the sign convention is $S^{(2)}=2\ln Z(\beta)-\ln Z(2\beta)$ and $\mathrm{DRTE}=\partial_gS^{(2)}$, as fixed by Eq.~\eqref{eq:RTEsignConvention}. The three transition classes are distinguished by the organization of the low-energy trace sampled by the two ensembles.

At a continuous transition, one critical tower controls the finite-size partition function:
\begin{equation}
    S^{(2)}=\widetilde S_2(x,c)+\cdots,
    \qquad
    \partial_gS^{(2)}
    =L^{1/\nu}\partial_x\widetilde S_2(x,c)+\cdots,
    \qquad
    x=(g-\gc)L^{1/\nu} .
    \label{eq:summaryContinuous}
\end{equation}
At fixed $c$, the leading RTE curves approach the common value $\widetilde S_2(0,c)$ at $g=\gc$, with finite-size drift generated by irrelevant fields and metric factors. The DRTE scales as $L^{1/\nu}$, but no thermodynamic identity requires either a zero at $\gc$ or a two-lobe profile. In the simulations, $c=\beta/L=1$, while the metric-normalized temporal ratio is $\rho_\tau=v$.

At a standard first-order transition, two frozen gapped branches dominate:
\begin{equation}
    \partial_gS_{\FO}^{(2)}
    =2\beta\partial_g\Delta E(g,L)
      [p_2(y)-p_1(y)],
    \qquad
    y=\beta\Delta E .
    \label{eq:summaryFOexact}
\end{equation}
Consequently,
\begin{equation}
    \partial_gS_{\FO}^{(2)}
    =L^{d+z_\beta}\mathcal F_2\!\left(
      [g-g_0(L)]L^{d+z_\beta};d_2/d_1
      \right)+\cdots .
    \label{eq:summaryFOFSS}
\end{equation}
Under the assumptions of Appendix~\ref{sec:firstorder}, the occupation identity yields two opposite-sign lobes, an exact zero at the finite-size branch-energy crossing $g_0(L)$, a peak scale $L^{d+z_\beta}$, and a width $L^{-(d+z_\beta)}$. The zero need not coincide with the thermodynamic $\gc$ until the crossing drift has vanished. The absolute RTE at coexistence is $\ln(d_1+d_2)$, rather than zero, unless a separate baseline is subtracted.

At a weakly first-order transition, coexistence is dressed by aspect-ratio-dependent pseudocritical towers. The general leading result is
\begin{equation}
\begin{split}
    \partial_gS^{(2)}
    =&\;2cL^{d+z_\beta}\Delta e'(g)
       [P_{2,2}-P_{2,1}]\\
     &+L^{1/\nu_\star}\mathcal B(x,c,L)+\cdots,
\end{split}
    \label{eq:summaryWFOgeneral}
\end{equation}
where
\begin{equation}
    \mathcal B
    =2\sum_{\alpha}P_{\alpha,1}\partial_x\ln Q_{\alpha,1}
     -\sum_{\alpha}P_{\alpha,2}\partial_x\ln Q_{\alpha,2} .
    \label{eq:summaryWFOtower}
\end{equation}
The first line is the macroscopic occupation-transfer term and naturally generates a sign-changing two-lobe component. The second contains the common critical-tower response and relative tower corrections. For $L\lesssim\xi_\star$, it may add a smooth peak or offset, shift the zero of the full DRTE, or dominate its magnitude.

Under the additional single-relative-mode condition of Eq.~\eqref{eq:WFOsingleMode},
\begin{equation}
\begin{split}
    \partial_gS^{(2)}
    =&\;L^{1/\nu_\star}\partial_xA_{\rm com}\\
     &+\left[
       2cL^{d+z_\beta}\Delta e'(g)
       +2cL^{1/\nu_\star}\phi'(x)
       \right]
       [P_{2,2}-P_{2,1}]+\cdots .
\end{split}
    \label{eq:summaryWFOreduced}
\end{equation}
Equation~\eqref{eq:WFOmainTextDictionary} gives the dictionary to the compact main-text formula. That expression applies when $\partial_xA_{\rm com}$ is negligible, forbidden by symmetry, or explicitly removed as a smooth background. In this regime, the occupation component has a robust zero, and its two-lobe structure can coexist with apparent $L^{1/\nu_\star}$ peak scaling. An exact zero of the full finite-size DRTE requires $\partial_xA_{\rm com}=0$ at the same point; a subleading but nonzero common-tower term shifts the zero by a finite-size correction.

These results sharpen the abbreviated statements in the main text. A collapse with an effective $\nu_\star$ signals an extended pseudocritical regime, but does not by itself distinguish a continuous transition from a walking first-order transition. By contrast, a reproducible opposite-sign component governed by occupation transfer between the $\beta$ and $2\beta$ ensembles provides evidence for competing spectral sectors. A generic smooth function can also display two extrema, so coexistence should be tested through the occupation identity, finite-size drift, aspect-ratio dependence, and robustness under smooth-background subtraction.

The spacetime Casimir interpretation is equally specific. Figure~\ref{fig:sc} compares one $2\beta$ cylinder with two disconnected $\beta$ cylinders. After ultraviolet-local terms linear in the temporal circumference cancel, the remaining $O(1)$ contribution is determined by the finite-aspect-ratio critical tower. The red curves schematically represent the change in infrared Matsubara spacing; they are neither exact many-body levels nor a mechanical force.

The formulas above assume that the low-temperature trace is dominated by the stated tower or branch structure. Corrections may arise from irrelevant fields, nonlinear scaling variables, a $g$-dependent velocity, finite-size crossing shifts, low-lying intrabranch modes, tunneling, interfaces, additional branches, or incomplete separation of common and relative tower sectors. Such effects modify drifts, offsets, and line shapes without changing the organizing hierarchy: a single critical tower, two frozen gapped branches, or two coexisting branches dressed by a long pseudocritical crossover.

\section{Application to the SU(8) $J_1$-$J_2$ model}
\label{app:SU(8)}
\begin{figure}[!h]
    \centering
    \includegraphics[width=0.7\linewidth]{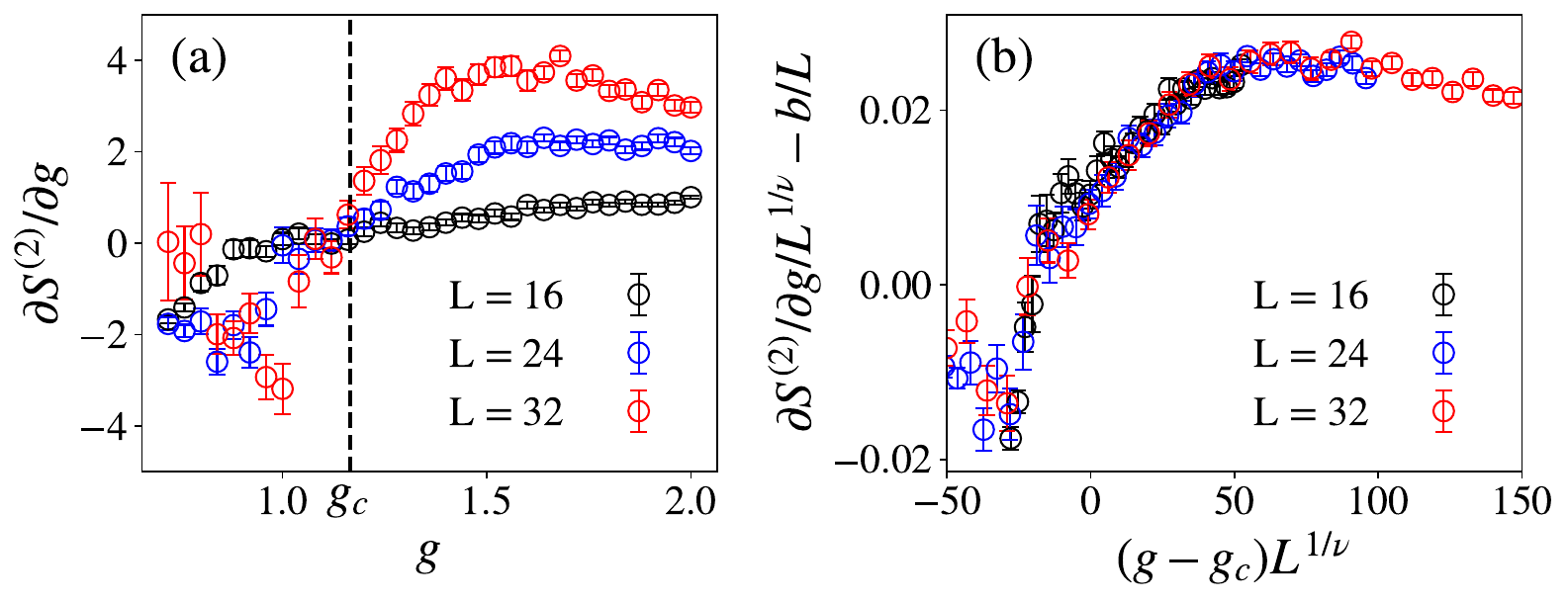}
    \caption{Application to the square-lattice SU(8) $J_1$-$J_2$ model. Here $g=J_2/J_1$, and the inverse temperature is fixed to $\beta=4L$; the two thermal ensembles entering the DRTE are therefore sampled at $\beta=4L$ and $2\beta=8L$. (a) DRTE $\partial S^{(2)}/\partial g$ for $L=16,24,32$ in the range $0.7\leq g\leq 2.0$. The vertical dashed line marks the fitted transition point, $g_c=1.17(4)$. The negative and positive lobes around $g_c$, together with the zero crossing, are the finite-size coexistence signature expected for a first-order transition. (b) Data collapse of the same data using $(g-g_c)L^{1/\nu}$ and $\partial S^{(2)} / \partial g / L^{1/\nu}-b/L$, with $g_c=1.17(4)$ and $\nu=0.67(4)$. The collapse captures the pseudocritical scaling regime.}
    \label{fig:su8}
\end{figure}

We further test the DRTE diagnostic in the square-lattice SU($N$) $J_1$-$J_2$-$Q$ model, which has been used in recent studies of SU($N$) deconfined quantum criticality~\cite{lou2009antiferromagnetic,kaul2012lattice,song2025evolution}. The model is defined on a bipartite square lattice, with SU($N$) spins in the fundamental representation on one sublattice and in the conjugate representation on the other. Its Hamiltonian is
\begin{equation}
H =
-\frac{J_1}{N}\sum_{\langle ij\rangle} P_{ij}
-\frac{J_2}{N}\sum_{\langle\langle ij\rangle\rangle} \Pi_{ij}
-\frac{Q}{N^2}\sum_{\langle i,j\rangle,\langle k,l\rangle} P_{ij}P_{kl}.
\label{eq}
\end{equation}
Here $P_{ij}$ projects a nearest-neighbor pair on opposite sublattices into an SU($N$) singlet, and $\Pi_{ij}$ permutes the colors of two next-nearest-neighbor sites on the same sublattice. The $J_1$ term favors antiferromagnetic correlations, the $J_2$ term favors Néel order for large enough $J_2/J_1$, and the $Q$ term is used mainly for smaller $N$ to stabilize the VBS phase. For $N>4$, the transition can be studied at $Q=0$ by tuning $g=J_2 / J_1$. In the following, we set the energy unit by $J_1=1$, take $N=8$ and $Q=0$, and tune $g$ across the transition.

\begin{figure*}[htb]
\centering
\subfigure[Binder cumulant of the $J$-$Q_2$ model]{
    \includegraphics[width=0.42\textwidth]{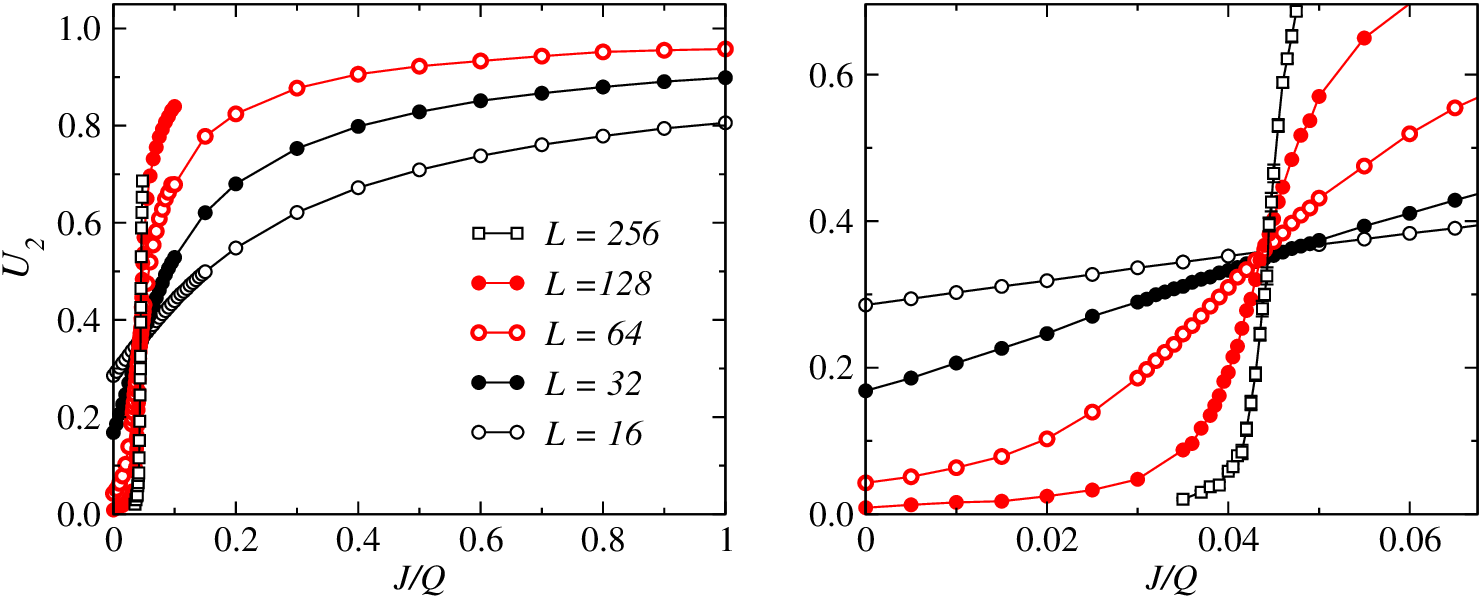}
    \label{jqbinder}
}

\subfigure[Spin and dimer correlations at distance $L/2$]{
    \includegraphics[width=0.60\textwidth]{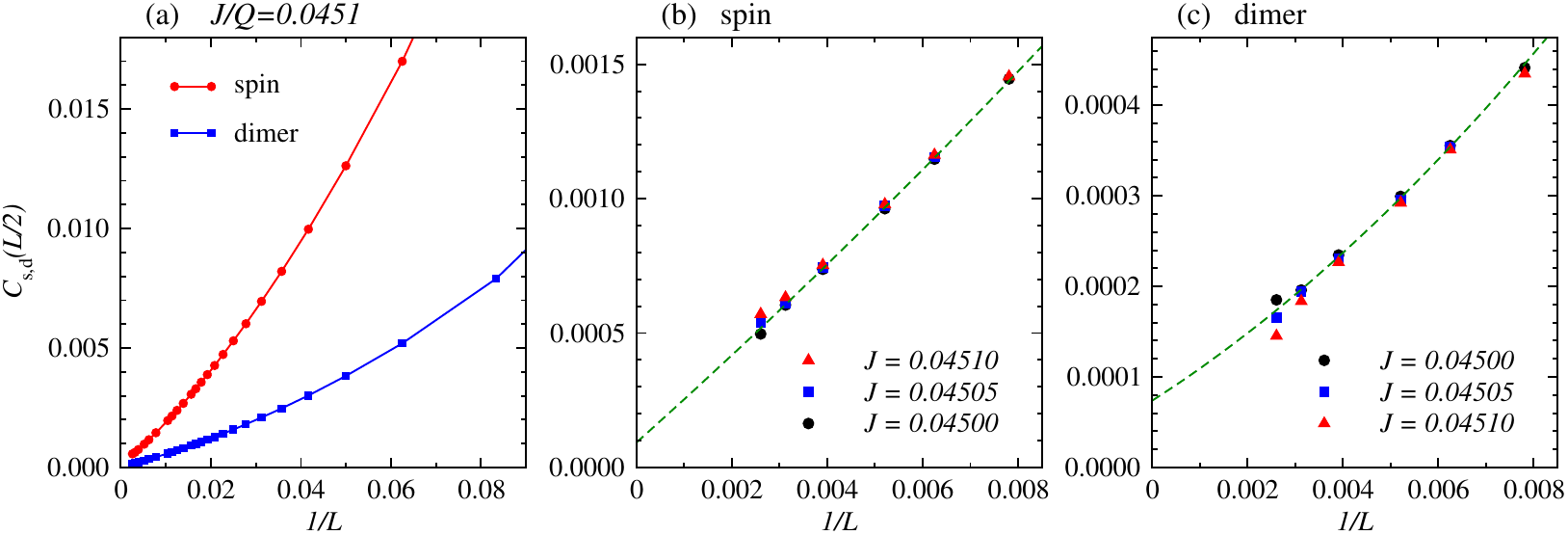}
    \label{jq2corre}
}

\caption{(a) Binder cumulant of the $J$-$Q_2$ model~\cite{sandvik2007evidence,sandvik2010computational}. The left panel shows results up to $J/Q=1$, where $U_2$ approaches $1$, indicating a N\'eel state. For smaller $J/Q$ the cumulant vanishes with increasing size, demonstrating a nonmagnetic state for $J/Q<0.045$. The cumulant remains positive for all $J/Q$, showing no phase coexistence, i.e., the transition is continuous based on these results.  (b) Spin and dimer correlation functions at distance $L/2$ vs the inverse system size in the critical $J$-$Q_2$ model, obtained with QMC calculations~\cite{Jun2024SO(5)}. Both correlation functions are shown vs $1/L$, with $L$ from $16$ (spin) or $12$ (dimer) up to $L=384$. The green dashed curves are cubic collective fits to the data. }
\label{fig:combined}
\end{figure*}

This model provides a useful test case because a recent entanglement entropy study suggested that the threshold value $N_c$ separating weaklyly first-order behavior from a continuous SU($N$) deconfined quantum critical point lies between $N=7$ and $N=8$~\cite{song2025evolution}. Thus SU(8) is close to the proposed boundary and corresponds to a continuous phase transition. We do not aim here to determine $N_c$ systematically. Instead, we use SU(8) as a focused test of whether the DRTE can detect a first-order component in a case where entanglement entropy scaling was interpreted as being compatible with a continuous transition.

The DRTE is obtained by evaluating this operator separately in the $\beta$ and $2\beta$ ensembles. In this calculation we use $\beta=4L$, so that the two ensembles entering DRTE are sampled at $\beta=4L$ and $2\beta=8L$. This choice is used to reduce residual finite-temperature effects in the subtraction. The data shown below are obtained for $L=16,24,32$ over the same coupling range $0.7\leq g\leq 2.0$.

Fig.~\ref{fig:su8}(a) shows the DRTE $\partial S^{(2)} / \partial g$ as a function of $g$. The curves show a sign-changing structure around the transition: DRTE develops a negative peak on the smaller-$g$ side, crosses zero near the transition point, and then forms a positive peak on the larger-$g$ side. This behavior is the characteristic shape of a  first-order transition in our protocol. It is different from the single-peak scaling form of a continuous transition. In Fig.~\ref{fig:su8}(b) we did the data collapse with the form
\begin{equation}
\partial S^{(2)} / \partial g / L^{1/\nu}= \mathcal{F}\left[(g-g_c)L^{1/\nu}\right] + b / L.
\label{eq}
\end{equation}
The collapse gives
\begin{equation}
g_c=1.17(4),\qquad \nu=0.67(4).
\end{equation}
In the present context, $\nu$ should be understood as an effective pseudocritical exponent in the accessible finite-size regime, rather than as independent evidence for a stable continuous fixed point. The diagnostic information comes from the combination of the collapse in Fig.~\ref{fig:su8}(b) and the sign-changing DRTE profile in Fig.~\ref{fig:su8}(a).

These results indicate that the transition in the SU(8) $J_1$-$J_2$ model still shows weaklyly first-order behavior. This differs from the interpretation based on smooth-boundary entanglement entropy scaling, where SU(8) was placed at or above the proposed onset of continuous behavior~\cite{song2025evolution}. The SU(8) data show that the DRTE can reveal a first-order component with modest system sizes in a case where the transition is difficult to settle using other finite-size probes.

\section{Binder cumulant and correlation functions for the $J$-$Q_2$ model}
\label{Binderandcorrelation}

To further illustrate the practical advantage of our method, we have added
a comparison with conventional observables for the $J$-$Q_2$ model (see
Fig.~\ref{fig:combined} ).  The Binder cumulant computed
up to $L=256$ shows no negative values that would indicate a first-order
transition.  The spin and dimer correlation functions measured at the
largest separation, even with system sizes up to $L=384$, do not display a
clear tendency to converge to finite constants in the thermodynamic limit,
making it impossible to unambiguously identify phase coexistence.  In
stark contrast, our DRTE already provides a definitive, qualitative
signature of the weakly first-order transition at system sizes as small as
$L=56$.  It is also worth noting that the computational cost of standard
QMC calculations scales as $\beta L^2 \sim L^3$.  This implies that even
with roughly 300 times the computational effort required for our $L=56$
DRTE result, the conventional diagnostics still cannot conclusively
characterize the nature of the transition, whereas our method does so
cleanly and with minimal computational expense.

\section{Tensor-network implementation of R\'enyi thermal entropy}
\label{Tensor}

\begin{figure}[!tb]
\centering
\includegraphics[width=0.6\textwidth]{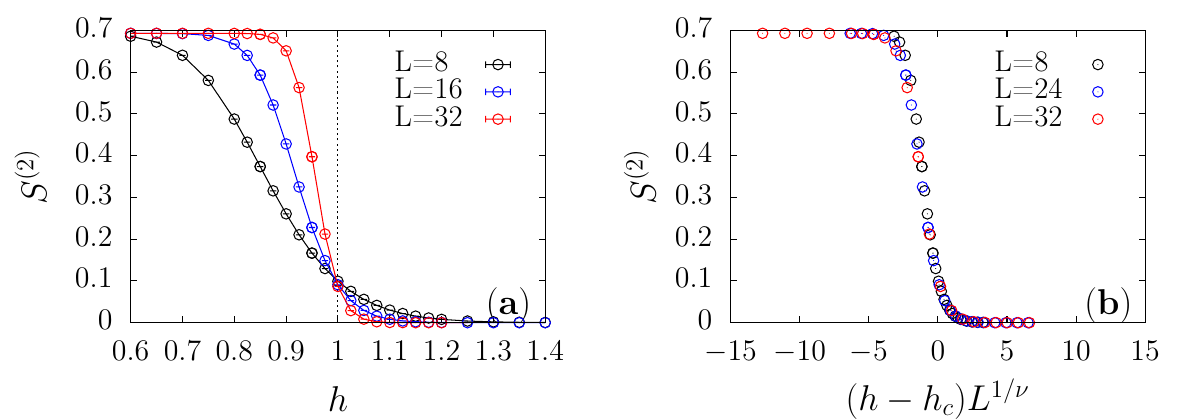}
\caption{(a) Second-order R\'enyi thermal entropy (RTE) and
(b) the corresponding data collapse near the quantum critical point
of the one-dimensional transverse-field Ising model
[$(1+1)$D Ising universality class].
The Hamiltonian is
$H=-J\sum_i\sigma_i^z\sigma_{i+1}^z-h\sum_i\sigma_i^x$,
with $J=1$ fixed and $h$ serving as the tuning parameter.
Applying the continuous-transition scaling form derived in the main text,
we obtain a critical field $h_c=0.99(1)$ and a critical exponent
$\nu=1.05(6)$, consistent within error bars with the exact results
$h_c=1$ and $\nu=1$.
All numerical results were obtained using the thermal tensor-network
method~\cite{Chen2018Exponential,Li2023Tangent}.
}
\label{fig:TH}
\end{figure}

The numerical results presented in the main text were obtained primarily
using quantum Monte Carlo simulations. Here we demonstrate that the
RTE framework is not restricted to a particular numerical method.
As a proof of principle, we implement the RTE using a thermal
tensor-network method for the one-dimensional transverse-field Ising
model,
\begin{equation}
H
=
-J\sum_i\sigma_i^z\sigma_{i+1}^z
-h\sum_i\sigma_i^x .
\label{eq:TN_TFIM}
\end{equation}
This model provides a particularly clean benchmark because its critical
point and critical exponent are exactly known,
\begin{equation}
h_c=J,
\qquad
\nu=1.
\end{equation}

A thermal tensor-network calculation provides access to the
finite-temperature density operator $e^{-\beta H}$ and hence to the
partition function
\begin{equation}
Z(\beta)=\Tr e^{-\beta H}.
\end{equation}
The R\'enyi thermal entropy can therefore be calculated directly from
\begin{equation}
S^{(n)}
=
\frac{1}{1-n}
\left[
\ln Z(n\beta)-n\ln Z(\beta)
\right],
\label{eq:TN_RTE}
\end{equation}
which is exactly the same quantity used in the quantum Monte Carlo
implementation discussed in the main text.

For the transverse-field Ising model, the dynamical exponent is $z=1$,
and we choose $\beta=L$. Near the quantum critical point, the RTE obeys
the continuous-transition scaling form
\begin{equation}
S^{(2)}
=
\widetilde S
\left[
(h-h_c)L^{1/\nu}
\right]
+\cdots .
\label{eq:TN_scaling}
\end{equation}
As shown in Fig.~\ref{fig:TH}(a), the RTE curves for different system
sizes exhibit a clear crossing near the quantum critical point.
Applying Eq.~\eqref{eq:TN_scaling}, we obtain
\begin{equation}
h_c=0.99(1),
\qquad
\nu=1.05(6),
\end{equation}
which agree with the exact results $h_c=1$ and $\nu=1$ within the
numerical uncertainty. The corresponding data collapse is shown in
Fig.~\ref{fig:TH}(b).

The DRTE can be evaluated within the same thermal tensor-network
framework. For a general tuning parameter $g$ in the Hamiltonian,
\begin{equation}
\frac{\partial S^{(n)}}{\partial g}
=
\frac{1}{1-n}
\left[
-n\beta
\left\langle
\frac{\partial H}{\partial g}
\right\rangle_{n\beta}
+
n\beta
\left\langle
\frac{\partial H}{\partial g}
\right\rangle_{\beta}
\right],
\label{eq:TN_DRTE}
\end{equation}
where the two expectation values are evaluated using the thermal
density matrices at inverse temperatures $n\beta$ and $\beta$,
respectively. Thus, both the RTE and DRTE can be implemented directly
within finite-temperature tensor-network calculations.

We emphasize that the transverse-field Ising example is intended as a
benchmark demonstrating the numerical portability of the RTE
framework, rather than as an independent tensor-network realization of
a weaklyly first-order transition. The weaklyly first-order theory developed
in the main text is formulated directly in terms of the finite-size
partition function and therefore does not depend on whether the
partition function is evaluated using quantum Monte Carlo, tensor
networks, or another numerical method.

In particular, when two competing thermodynamic branches contribute
appreciably to the finite-size partition function,
\begin{equation}
Z(s\beta)
\simeq
Z_1(s\beta)+Z_2(s\beta),
\label{eq:TN_two_branch}
\end{equation}
the DRTE probes how their relative statistical weights differ between
the $\beta$ and $n\beta$ ensembles. Therefore, the same two-branch
mechanism derived in the main text predicts a zero crossing and an
opposite-sign double-peak structure in a tensor-network calculation of
a weaklyly first-order transition.

The advantage of the RTE framework emerges once the system enters
the low-energy regime. The R'enyi combination
\begin{equation}
S^{(n)}
=
\frac{1}{1-n}
\left[
\ln Z(n\beta)-n\ln Z(\beta)
\right]
\end{equation}
cancels the dominant smooth contribution that is linear in the
imaginary-time extent. Consequently, the universal or
coexistence-controlled finite-size contribution becomes more directly
visible. This mechanism is independent of the numerical technique used
to evaluate the partition functions.

Therefore, the transverse-field Ising calculation presented here serves
as a controlled benchmark showing that the RTE finite-size scaling
framework can be implemented using thermal tensor networks and correctly
reproduces known critical properties. The same partition-function-based
construction and weaklyly first-order scaling theory can be applied to
tensor-network calculations of more general quantum many-body systems.

\end{widetext}
\end{document}